\documentclass[usenatbib]{mnras}
\defcitealias{Sironi_20}{SB20}

\usepackage[T1]{fontenc}
\usepackage{ae,aecompl}

\usepackage{graphicx}	% Including figure files
\usepackage{amsmath}	% Advanced maths commands
\usepackage{amssymb}	% Extra maths symbols
\usepackage{threeparttable}
\usepackage{booktabs,caption}
\usepackage{changepage}
\usepackage{adjustbox}
\usepackage{graphics}
\usepackage{subcaption}
\usepackage{mathrsfs}
\usepackage{lipsum,verbatim}
\usepackage{ulem}
\newcommand{\ux}{u_{\rm x}}
\newcommand{\uy}{u_{\rm y}}
\newcommand{\Eint}{{\cal E}_{\rm int}}
\newcommand{\gcr}{\gamma_{\rm cr}}
\newcommand{\gammacr}{\gamma_{\rm cr}}
\newcommand{\gammae}{\gamma_{\rm e}}
\newcommand{\comp}{c/\omega_{\rm p}}

\newcommand{\fig}[1]{Fig.~\ref{fig:#1}}
\newcommand{\gcool}{\tau_{\rm cool}}

\newcommand{\eq}[1]{Eq.~(\ref{eq:#1})}
\newcommand{\fign}[1]{\ref{fig:#1}}

\newcommand{\eqb}{\begin{eqnarray}}
\newcommand{\eqe}{\end{eqnarray}}
\newcommand{\be}{\begin{eqnarray}}
\newcommand{\ee}{\end{eqnarray}}
\newcommand{\bi}{\begin{itemize}}
\newcommand{\ei}{\end{itemize}}
\newcommand{\fHE}{f_{\rm HE}}

\title[Cold-chain Comptonization in black hole coronae I]{Comptonization by Reconnection Plasmoids in Black Hole Coronae I: Magnetically Dominated Pair Plasma}

\author[N. Sridhar, L. Sironi \& A. Beloborodov]{
Navin Sridhar,$^1$\thanks{E-mail: navin.sridhar@columbia.edu}
Lorenzo Sironi,$^1$\thanks{E-mail: lsironi@astro.columbia.edu} and
Andrei M. Beloborodov$^{2,3}$\thanks{E-mail: amb2046@columbia.edu }
\\
$^1$Department of Astronomy and Columbia Astrophysics Laboratory, Columbia University, 550 W 120th St, New York, NY 10027, USA\\
$^2$Department of Physics and Columbia Astrophysics Laboratory, Columbia University, 550 W 120th St, New York, NY 10027, USA\\
$^3$Max Planck Institute for Astrophysics, Karl-Schwarzschild-Str. 1, D-85741, Garching, Germany
}

\date{Accepted XXX. Received YYY; in original form ZZZ}

\pubyear{2021}

\begin{document}
\label{firstpage}
\pagerange{\pageref{firstpage}--\pageref{lastpage}}
\maketitle

% Abstract of the paper
\begin{abstract}
We perform two-dimensional particle-in-cell simulations of reconnection in magnetically dominated electron-positron plasmas subject to strong Compton cooling. We vary the magnetization $\sigma\gg1$, defined as the ratio of magnetic tension to plasma inertia, and the strength of cooling losses. Magnetic reconnection under such conditions can operate in magnetically dominated coronae around accreting black holes, which produce hard X-rays through Comptonization of seed soft photons. We find that the particle energy spectrum is dominated by a peak at mildly relativistic energies, which results from bulk motions of cooled plasmoids. The peak has a quasi-Maxwellian shape with an effective temperature of $\sim 100$~keV, which depends only weakly on the flow magnetization and the strength of radiative cooling. The mean bulk energy of the reconnected plasma is roughly independent of $\sigma$, whereas the variance is larger for higher magnetizations. The spectra also display a high-energy tail, which receives $\sim 25$\% of the dissipated reconnection power for $\sigma=10$ and $\sim 40$\% for $\sigma=40$. We complement our particle-in-cell studies with a Monte-Carlo simulation of the transfer of seed soft photons through the reconnection layer, and find the escaping X-ray spectrum. The simulation demonstrates that Comptonization is dominated by the bulk motions in the chain of Compton-cooled plasmoids and, for $\sigma\sim 10$, yields a spectrum consistent with the typical hard state of accreting black holes.
\end{abstract}

% Select between one and six entries from the list of approved keywords.
% Don't make up new ones.

%%%%%%%%%%%%%%%%%%%%%%%%%%%%%%%%%%%%%%%%%%%%%%%%%%

%%%%%%%%%%%%%%%%% BODY OF PAPER %%%%%%%%%%%%%%%%%%

\section{Introduction}

The emission mechanism of high-energy non-thermal X-rays from black hole X-ray binaries (BHXBs) is still unknown. Non-thermal X-rays are predominantly seen during the so-called ``hard state''---typically observed during the onset as well as the late-time fading of an outburst. During outbursts, the X-ray luminosity increases by a few orders of magnitude, with changes in the radiation spectrum \citep{2006csxs.book..157M}, light curve variability \citep{1989ARA&A..27..517V}, and possible launching of transient radio jets and collimated outflows \citep{1999ARA&A..37..409M, 2004MNRAS.355.1105F, 2009MNRAS.396.1370F}. An example of the transitions across different states of an outburst, and their physical origins, in the archetypal BHXB GX~339--4, is demonstrated in \cite{Sridhar+20}. 

A typical photon spectrum during the hard state may be roughly described as ${\rm d}N/{\rm d}E\propto E^{-\Gamma}$ with index $\Gamma\lesssim1.8$ (where $E$ is the photon energy) and an exponential cutoff above $\sim 100$~keV. The hard X-ray emission is commonly attributed---based on the quality of spectral fits---to unsaturated Comptonization of soft photons by a cloud of hot electrons called ``corona,'' with typical temperature of $\sim 100$ keV \citep{Zdziarski&Gierlinski_04}.

\begin{figure} 
\fboxsep=2pt  %padding thickness
\fboxrule=1pt %border thickness
\fcolorbox{black}{white}{\includegraphics[width=8.0cm]{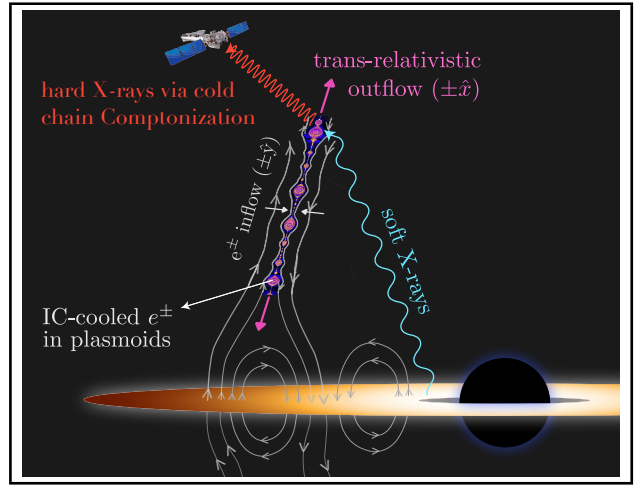}}
\caption{A schematic of the ``chain Comptonization'' model. The black sphere embedded in a golden-brown disk represents the black hole-accretion disk system. Differential rotation of the magnetic field footpoints in the accretion disk leads to stretching and opening of the field lines. Two oppositely oriented fields are separated by a current sheet, and the energy in the magnetic loop is released via reconnection. This process heats the coronal particles, while the soft disk photons (blue) cool them to non-relativistic temperatures via inverse Compton (IC) scattering. Reconnection also generates a chain of coherent magnetic structures called ``plasmoids'' (magenta blobs), which get accelerated  to trans-relativistic speeds along the layer by magnetic tension. Comptonization of soft disk photons by the bulk motions of a cold-chain of plasmoids can reproduce the non-thermal/hard X-ray emission (red) observed from X-ray binaries. 
}
\label{fig:illustration}
\end{figure}

Yet, the energization mechanism that sustains the coronal electrons against fast inverse Compton (IC) losses is still unclear. A number of works invoked magnetic reconnection as a mechanism for heating and acceleration of electrons in black hole coronae \citep[e.g.,][]{Galeev+79, Beloborodov_99, liu_02}. Numerical simulations demonstrate that current sheets can form as magnetic loops get twisted by the differential rotation of the accretion flow  and the black hole \citep{parfrey_15, Yuan+19, ripperda_20, Krawczynski_20, Chashkina+21}. Figure~\ref{fig:illustration} illustrates one such possible configuration. Fast magnetic reconnection (``relativistic regime'') can occur in the current sheets above the accretion disk, where the energy density in magnetic fields $B^2/8\pi$ exceeds the plasma rest mass energy density $\rho c^2$, which corresponds to magnetization parameter $\sigma\equiv B^2/4\pi\rho c^2 > 1$ \citep[for reviews of relativistic reconnection, see, e.g.,][]{hoshino_lyubarsky_12,kagan_15}. Kinetic particle-in-cell (PIC) simulations show how relativistic reconnection heats and accelerates plasma particles. Most PIC simulations have been conducted in the regime of negligible radiative losses \citep[e.g.,][]{zenitani_01, lyubarsky_liverts_08, kagan_13, sironi_spitkovsky_14, guo_14, guo_19, nalewajko_15, werner_16, werner_17, sironi_15, sironi_16, petropoulou_18,hakobyan_20,zhang_21}. 

Some recent studies have incorporated IC cooling effects, e.g., \cite{nalewajko_18}, \cite{werner_19}, \cite{Sironi_20} (\citetalias{Sironi_20}, hereafter), and \cite{Mehlhaff_20}. In all these simulations, magnetic reconnection is developed through fragmentation of the dissipation layer into a chain of numerous plasmoids (magnetic islands), which move at relativistic speeds along the layer, as predicted by analytical models \citep{uzdensky_10}.
 
Radiative cooling of electrons in luminous sources (e.g., Cygnus~X-1), is very fast---the cooling timescale is much shorter than the light-crossing time of the corona. \citet{belo_17} pointed out that in this regime most of the plasma in the reconnection layer is kept at the local Compton temperature $kT_{\rm C}\ll 100\,$keV, and Comptonization of hard X-rays mainly results from the fast bulk motions of the cold plasmoid chain accelerated along the layer by the tension of magnetic field lines. Radiation exerts resistance to the plasmoid motion, as if they moved through a viscous background, and magnetic energy is passed through the plasmoids directly to photons, with subdominant heating of individual particles. Monte-Carlo simulations of this Comptonization mechanism suggested an intriguing feature: it naturally gives an X-ray spectrum peaking at $\sim 100\,$keV, consistent with the observed hard-state spectra of accreting black holes in BHXBs and Active Galactic Nuclei (AGN).
 
This mechanism was further demonstrated by the kinetic plasma simulations of \citetalias{Sironi_20}. They found that 70-80\% of the reconnection power converts to radiation via Comptonization by the plasmoid bulk motions, and that these motions mimic a quasi-thermal distribution with $kT_{\rm e}\sim 100\,$keV. Their simulations have been performed for $e^\pm$ plasma with magnetization $\sigma=10$. The dependence on $\sigma$ and the possible role of ions remained unexplored.

In the present paper, we extend the PIC simulations of \citetalias{Sironi_20}---with strong radiative (IC) cooling---to a higher $\sigma=40$ and quantify the dependence of particle heating and acceleration, as well as of reconnection-induced bulk motions, on the flow magnetization and the strength of cooling. In addition, we evaluate the produced X-ray spectrum with Monte-Carlo simulations for $\sigma=10$ and 40.

Like \citetalias{Sironi_20}, our PIC simulations are performed for $e^\pm$ plasma. We assume that the $e^\pm$ are created by the MeV photons in the tail of the Comptonized radiation spectrum, which requires a sufficiently high compactness parameter of the magnetic flare \citep{belo_17}.

Ion density in the high-$\sigma$ coronae of black holes is unknown. Since ions are not subject to radiative losses, they could store part of the released magnetic energy and gradually transfer it to the electrons. This could influence the emission mechanism of the magnetic flare. As a first step toward understanding the possible effect of ions, we perform an experiment where  positrons play the role of ``ions'' with mass $m_{\rm i}=m_{\rm e}$, and only electrons are subject to radiative losses. More realistic simulations with $m_{\rm i}\gg m_{\rm e}$ are more expensive and deferred to part II of this series.

The paper is organized as follows. In \S\ref{sec:setup}, we describe the numerical setup of our simulations. In \S\ref{sec:IC_cooling}, we describe the implementation and parameterization of IC cooling, and the different timescales associated with the problem. In \S\ref{sec:results}, we present our PIC results, emphasizing the dependence on magnetization and strength of IC cooling. The photon spectra derived from our PIC simulations using Monte Carlo radiative transfer calculations are presented in \S\ref{sec:radiative_transfer}. We summarize our findings in \S\ref{sec:conclusion}.

%%%%%%%%%%%%%%%%%%%%%%%%%%%
\section{PIC simulation setup} \label{sec:setup}
The simulations are performed with the 3D electromagnetic PIC code \textsc{tristan-mp} \citep{buneman_93, spitkovsky_05}, with particle momentum updated via the Vay pusher \citep{Vay_08}. We employ a 2D spatial domain in the $x$-$y$ plane, but all three components of velocities and  electromagnetic fields are evolved. The simulation setup parallels what we have employed in, e.g., \citealt{sironi_16} and \citetalias{Sironi_20}. We refer to Table \ref{tab:setup} in Appendix \ref{appendix:table} for the complete  set of our numerical and physical input parameters.

We consider a plasma consisting of two particle species of equal mass and opposite charge (electrons and positrons). The particles are initialized with a uniform density $n_{\rm 0}$ and a small thermal spread $kT_{0}/m_{\rm e}c^2=10^{-4}$. The reconnection layer is set up with the so-called Harris equilibrium \citep{1962NCim...23..115H}, where the initial magnetic field is $B_x=-B_{\rm 0}\tanh(2\pi y/\Delta)$, i.e., the field reverses at $y=0$ across a thickness $\Delta$. We also add a uniform magnetic field aligned with the electric current (i.e., a guide field $B_{\rm g}=0.1\,B_{\rm 0}$ along $z$). This helps providing  pressure support to the cores of plasmoids, which get significantly compressed due to cooling. Note that the addition of the guide field does not add free energy to the system, as it does not get dissipated. 

The field strength is parameterized by the plasma magnetization, $\sigma$. For a cold plasma---as it is in our case---we define the magnetization as 
\begin{equation} \label{eqn:magnetization}
\sigma=\dfrac{B^2_{\rm 0}}{4\pi n_{\rm 0}m_{\rm e}c^2} = \bigg(\dfrac{\omega_{\rm c}}{\omega_{\rm p}}\bigg)^2,
\end{equation}
where $\omega_{\rm c}=eB_{\rm 0}/m_{\rm e}c$ is the electron gyro-frequency, and the plasma frequency $\omega_{\rm p}$ is defined as

\begin{equation} \label{eqn:plasma_freq}
\omega_{\rm p}=\sqrt{\frac{4\pi n_{\rm 0} e^2}{m_{\rm e}}}~.
\end{equation}
The corresponding plasma skin depth is $c/\omega_{\rm p}$. We are interested in the regime of relativistic reconnection $\sigma\gg1$---specifically, we consider $\sigma=40$ (our fiducial case) and $\sigma=10$ (\citetalias{Sironi_20}'s reference case). The  Alfv\'{e}n speed is $v_{\rm A}/c=\sqrt{\sigma/(\sigma+1)}\simeq 1$, i.e., close to the speed of light.

Initially, magnetic pressure outside the sheet is balanced by thermal pressure in the sheet, by adding a hot population with over-density $\eta=3$ relative to the number density $n_{\rm 0}$ of cold particles outside the sheet. The hot population has a temperature of $k_{\rm B}T_{\rm h}/m_{\rm e}c^2=\sigma/2\eta$. Reconnection is triggered by reducing the temperature of the hot particles near the center of the domain [$(x,y)=(0,0)$] at the initial time. This results in the formation of a magnetic ``X-point'' at the center of the computational domain. The plasma in the current sheet separates into two reconnection fronts, that propagate away from the center along $\pm\hat{x}$ at near the Alfv\'{e}n speed. The initial current sheet thickness is chosen to be large enough ($\Delta \simeq 30~\sqrt{\sigma}\,c/\omega_{\rm p}$), so that the tearing mode does not spontaneously grow before the two reconnection fronts have reached the boundaries of the domain.

We employ outflow boundary conditions along $x$ \citep[e.g.,][]{sironi_16}, so the hot particles initialized in the sheet get evacuated after $\sim L_{\rm x}/v_{\rm A}$, where $L_{\rm x}$ is the half-length of the box along $x$. Along the $y$ direction of the reconnection inflow, we employ two moving injectors---receding from $y = 0$ at the speed of light and continuously introducing fresh plasma and magnetic flux into the domain---and an expanding simulation box \citep[see][for details]{sironi_16}. The combination of outflow boundaries in $x$ and continuous injection in $y$ ensures that, after $\sim1$ Alfv\'{e}nic crossing time (see Appendix~\ref{appendix:time_convergence} for details), reconnection proceeds in a quasi-steady state, which is independent from the sheet initialization (i.e., from the choices of $\eta$ and $\Delta$), as shown by \cite{sironi_16}. We follow the evolution until $\sim 5 L_{\rm x}/c$, such that we have a sufficiently long time to assess the steady-state properties of the system.

A large dynamic range between plasma scales and the layer length is essential to obtain astrophysically-relevant results. Our reference box for $\sigma=40$ has $L_{\rm x}/(c/\omega_{\rm p})=3360$, but we also present results from a wide range of box sizes, $420\le L_{\rm x}/(c/\omega_{\rm p})\le 6720$. We resolve the plasma skin depth $c/\omega_{\rm p}$ with 5 grid lengths. The Courant-Friedrichs-Lewy number (or equivalently, the numerical speed of light) is set to 0.45.  We employ 4 particles per cell (including both species), and we improve particle noise in the electric current density with 32 passes of a ``1-2-1'' low-pass digital filter applied at each step \citep{birdsall_91}.

%%%%%%%%%%%%%%%%%%%%%%%%%%%%%%
\section{Inverse-Compton cooling} \label{sec:IC_cooling}

Compton cooling is implemented in our code as a ``drag'' force applied to the simulation particles \citep{2010NJPh...12l3005T}. For an electron (or positron) with velocity $\mathbf{v_{\rm e}}$ ($=\boldsymbol{\beta}_{\rm e}c$) and energy $\gamma_{\rm e} m_{\rm e}c^2$, the Compton drag force due to an isotropic distribution of photons is given by
\begin{equation} \label{eqn:f_ic}
{\mathbfit F}_{\rm IC} = -\frac{4}{3}\sigma_{\rm T}\gamma_{\rm e}^2U_{\rm rad} \boldsymbol{\beta}_{\rm e},
\end{equation} 
where $\sigma_{\rm T}=8\pi e^4/(3m_{\rm e}^2 c^4)$ is the Thomson cross section, $\gamma_{\rm e}=(1-\beta_{\rm e}^2)^{-1/2}$ is the particle Lorentz factor, and $U_{\rm rad}$ is the radiation energy density.
We parameterize $U_{\rm rad}$ by defining a critical Lorentz factor $\gamma_{\rm cr}$, at which the Compton drag force balances the force due to the reconnection electric field $E_{\rm rec}=\eta_{\rm rec}B_{\rm 0}$, where $\eta_{\rm rec}\sim0.1$ is the reconnection rate. For $|\beta_{\rm e}|\simeq1$, the balance
\begin{equation} \label{eqn:gamma_cr}
eE_{\rm rec} = \frac{4}{3}\sigma_{\rm T}\gamma_{\rm cr}^2U_{\rm rad}
\end{equation}
yields $\gamma_{\rm cr} \equiv \sqrt{3e\eta_{\rm rec}B_{\rm 0}/(4\sigma_{\rm T}U_{\rm rad})}$. A low value of $\gamma_{\rm cr}$ implies strong cooling (i.e., large $U_{\rm rad}$). In contrast,  the limit of negligible cooling losses ($U_{\rm rad}=0$) corresponds to $\gamma_{\rm cr} =\infty$. For our reference magnetization $\sigma=40$, we investigate $\gcr=16, 22.6, 32, 45$ and the uncooled case $\gcr=\infty$.

The IC cooling time for a particle with Lorentz factor $\gamma_{\rm e}$ can be written as
\begin{equation} \label{eqn:t_ic}
 t_{\rm IC}(\gamma_{\rm e})=\frac{3m_{\rm e}c}{4\sigma_{\rm T}U_{\rm rad}\gamma_{\rm e}}=\frac{1}{\omega_{\rm c}}\frac{\gamma_{\rm cr}^2}{\eta_{\rm rec}\gamma_{\rm e}}.
 \end{equation} 
This should be compared with the temporal resolution of our simulations
\begin{equation} \label{eqn:t_res}
\Delta t=0.09~\omega_{\rm p}^{-1}=0.09\,\sigma^{1/2}\omega_{\rm c}^{-1}.
\end{equation}
This shows that the IC cooling time for a particle with Lorentz factor as high as $\gamma_{\rm e}=\gamma_{\rm cr}$ is well resolved if 
\begin{equation} 
\frac{\Delta t}{t_{\rm IC}(\gamma_{\rm cr})} \sim 10^{-2}\frac{\sigma^{1/2}}{\gamma_{\rm cr}} \ll 1~.
\end{equation}
This condition is well satisfied in our simulations, which all have 
$\gamma_{\rm cr}> \sqrt{\sigma}$ (see Table \ref{tab:setup} in Appendix~\ref{appendix:table}).

%%%%%%%%%%
\subsection{Energy and time scales} \label{subsec:time_scales}

Here, we summarize the hierarchy of energy and time scales of radiative reconnection in BHXB coronae. We refer to \citet{belo_17} and \citetalias{Sironi_20} for additional details. 

As long as the Thomson optical depth of the reconnection layer is not much greater than unity ($\tau_{\rm T}\sim 1$ is typically inferred for black hole coronae\footnote{The optical depth across the layer is $\tau_{\rm T}=Hn_{\rm 0}\sigma_{\rm T}$, where $H\sim \eta_{\rm rec}L_{\rm x}$ is the thickness of the reconnection layer.}), energy conservation implies that the radiation density is $U_{\rm rad}\sim \eta_{\rm rec} U_{\rm B}\sim 0.1\,U_{\rm B}$, where $U_{\rm B}=B_{\rm 0}^2/8\pi$ is the energy density of the reconnecting field. The magnetization parameter $\sigma$ can be expressed as
\be\label{eq:sigma}
\sigma=\frac{2 U_{\rm B}}{n_{\rm 0} m_{\rm e} c^2}\sim \frac{2\eta_{\rm rec}\ell_{\rm B}}{\tau_{\rm T}}
\ee
where 
$\ell_{\rm B}=U_{\rm B} \sigma_{\rm T} L_{\rm x}/m_{\rm e} c^2$ is the magnetic compactness. For magnetic flares near black holes accreting at a significant fraction of the Eddington limit, the magnetic compactness  can approach $\ell_{\rm B}\sim m_{\rm p}/m_{\rm e}$ \citep{belo_17}. Using 
$\eta_{\rm rec}\sim 0.1$ and $\tau_{\rm T}\sim 1$, \eq{sigma} gives $\sigma\sim 200\,(\ell_{\rm B}/10^3)$. 

Given that $U_{\rm rad}\sim \eta_{\rm rec} U_{\rm B}$, we can quantify the expected value of $\gammacr$ as
\be
\gammacr=\left(\frac{27 \,L_{\rm x}}{16\, \ell_{\rm B} r_{\rm e}}\right)^{1/4}\sim 10^4 \left(\frac{M_{\rm BH}}{10M_\odot}\right)^{1/4}
\ee
where $r_{\rm e}$ is the classical electron radius, $\ell_{\rm B}\sim m_{\rm p}/m_{\rm e}$ and we have assumed that the characteristic length of reconnection layers is $L_{\rm x}\sim r_{\rm g}$, where $r_{\rm g}=2 G M_{\rm BH}/c^2$ is the Schwarzschild radius for a black hole of mass $M_{\rm BH}$.

Accelerated particles typically attain a Lorentz factor $\gamma_{\rm X}\sim \sigma/4$ at X-points. A high value of $\gammacr\gg \gamma_{\rm X}\sim \sigma/4$ then implies that particle acceleration at X-points is not impeded by Compton drag. The same condition can be expressed by comparing the IC cooling timescale for particles with $\gammae=\gamma_{\rm X}$ with the timescale $t_{\rm X}$ for particle acceleration at X-points,
\be
  \frac{t_{\rm IC}(\gamma_{\rm X})}{t_{\rm X}}=\frac{\gammacr^2}{\gamma_{\rm X}^2}\gg 1,
  \qquad t_{\rm X} =\frac{\gamma_{\rm X} m_{\rm e} c}{e E_{\rm rec}}~.
\ee

On the other hand, the IC cooling time is much shorter than the advection time along the layer $t_{\rm adv}\sim L_{\rm x}/v_{\rm A}\sim L_{\rm x}/c$. Their ratio is
\be
\frac{t_{\rm IC}(\gamma_{\rm e})}{t_{\rm adv}}=\frac{3}{4 \gamma_{\rm e} \ell_{\rm rad}} \ll 1
\ee
where the radiation compactness is defined as $\ell_{\rm rad}=U_{\rm rad} \sigma_{\rm T} L_{\rm x}/m_{\rm e} c^2=\eta_{\rm rec}\ell_{\rm B}$. We rewrite the ratio of timescales as
\begin{equation}
   \frac{t_{\rm IC}(\gamma_{\rm e})}{t_{\rm adv}}=\frac{\gcool}{\gamma_e}.
\end{equation}
Then, the condition that the time to cool down to non-relativistic energies is shorter than the advection timescale may be expressed as
\begin{equation} \label{eq:varrho}
\gcool \equiv \frac{\gcr^2}{\eta_{\rm rec}\sqrt{\sigma}}\frac{c/\omega_{\rm p}}{L_{\rm x}} < 1~.
\end{equation}
In our simulations, we explore both $\gcool<1$ and $\gcool>1$ regimes. In the latter case, $\gcool$ equals the Lorentz factor of particles that cool in a dynamical (advection) time. We use $\gcool$ (as an alternative to $\gcr$) to quantify the strength of cooling losses.

In the limit of strong cooling, particles in reconnection plasmoids are cooled down to non-relativistic temperatures ($\gamma_{\rm e}\approx 1$), well before the plasmoid bulk motions are decelerated by Compton drag \citep{belo_17}. For the cold particles, the ratio of drag time to advection time is 
\begin{equation} \label{eq:t_drag}
    \frac{t_{\rm drag}}{t_{\rm adv}} \sim \frac{\sigma m_{\rm e}c}{U_{\rm rad}\sigma_{\rm T}} \bigg/ \frac{L_{\rm x}}{c} \sim \frac{\sigma}{\ell_{\rm rad}}.
\end{equation}
Given that $\sigma\sim \eta_{\rm rec}\ell_{\rm B}/\tau_{\rm T}\sim \ell_{\rm rad}/\tau_{\rm T}$ (Eq. \ref{eq:sigma}), one can see that the two timescales are comparable for $\tau_{\rm T}\sim 1$. Their ratio may also be written as $t_{\rm drag}/t_{\rm adv}\sim \sigma\gcool$ when $\gamma_{\rm e}\approx 1$.

In summary, in order to mimic the conditions in black hole coronae, our simulations have to retain the following hierarchy of energy and time scales: ({\it i}) $\sigma\gg1$; ({\it ii}) $\gammacr\gg \gamma_{\rm X}\sim \sigma/4$, or equivalently particle acceleration at X-points occurs faster than IC cooling losses;
({\it iii}) $\gcool\ll1$, or equivalently the IC cooling time for all particles is shorter than the advection time along the layer. Most of our simulations do satisfy the required hierarchy.

%%%%%%%%%%%%
\section{PIC simulation results} \label{sec:results}

In this section, we present the main results of our PIC simulations. We discuss how our PIC results depend on the strength of radiative IC losses, magnetization and domain size. Our fiducial runs have $\sigma=40$ and $L_{\rm x}/(c/\omega_{\rm p})=3360$. 

In \S\ref{subsec:general}, we describe the general structure of the reconnection layer. We then explore the influence of radiative cooling on internal and bulk motions in \S\ref{subsec:velocities}. Particle energy spectra are presented in \S \ref{subsec:spectra}. In \S\ref{subsec:posicool}, we probe the structure of the reconnection layer and the particle energy spectrum in a ``hybrid'' experiment, where we consider a pair plasma, but only cool electrons. This is a first step towards understanding radiative reconnection in an electron-ion plasma, where only electrons suffer IC cooling losses. 

%%%%%%%%%%%%%%%%%%%%%
\begin{figure*}
    \centering 
    \begin{subfigure}[t]{0.5\textwidth}
        \centering
        \includegraphics[height=2.8in]{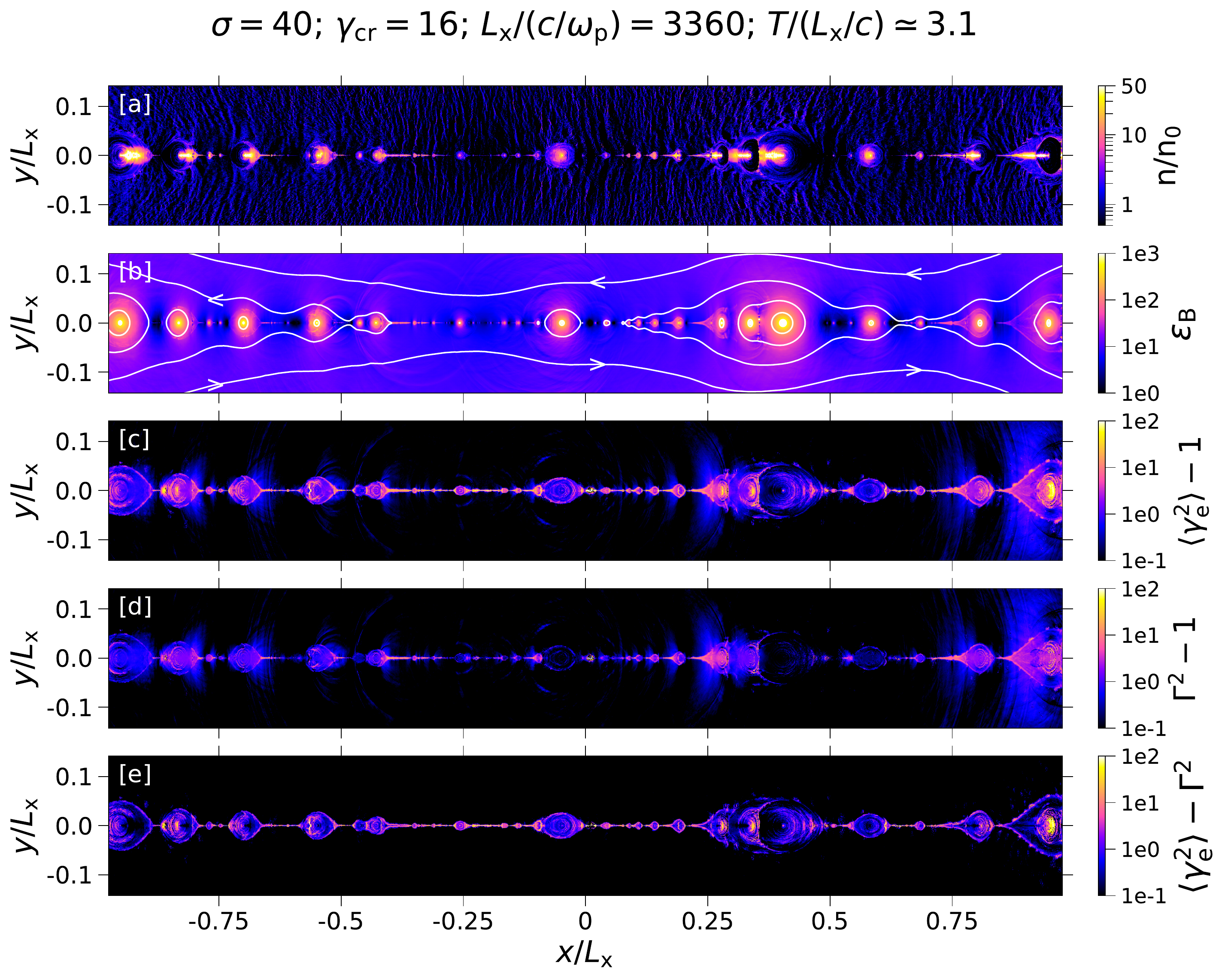}
        \caption*{(i) Strong IC cooling}
    \end{subfigure}%
    ~ 
    \begin{subfigure}[t]{0.5\textwidth}
        \centering
        \includegraphics[height=2.8in]{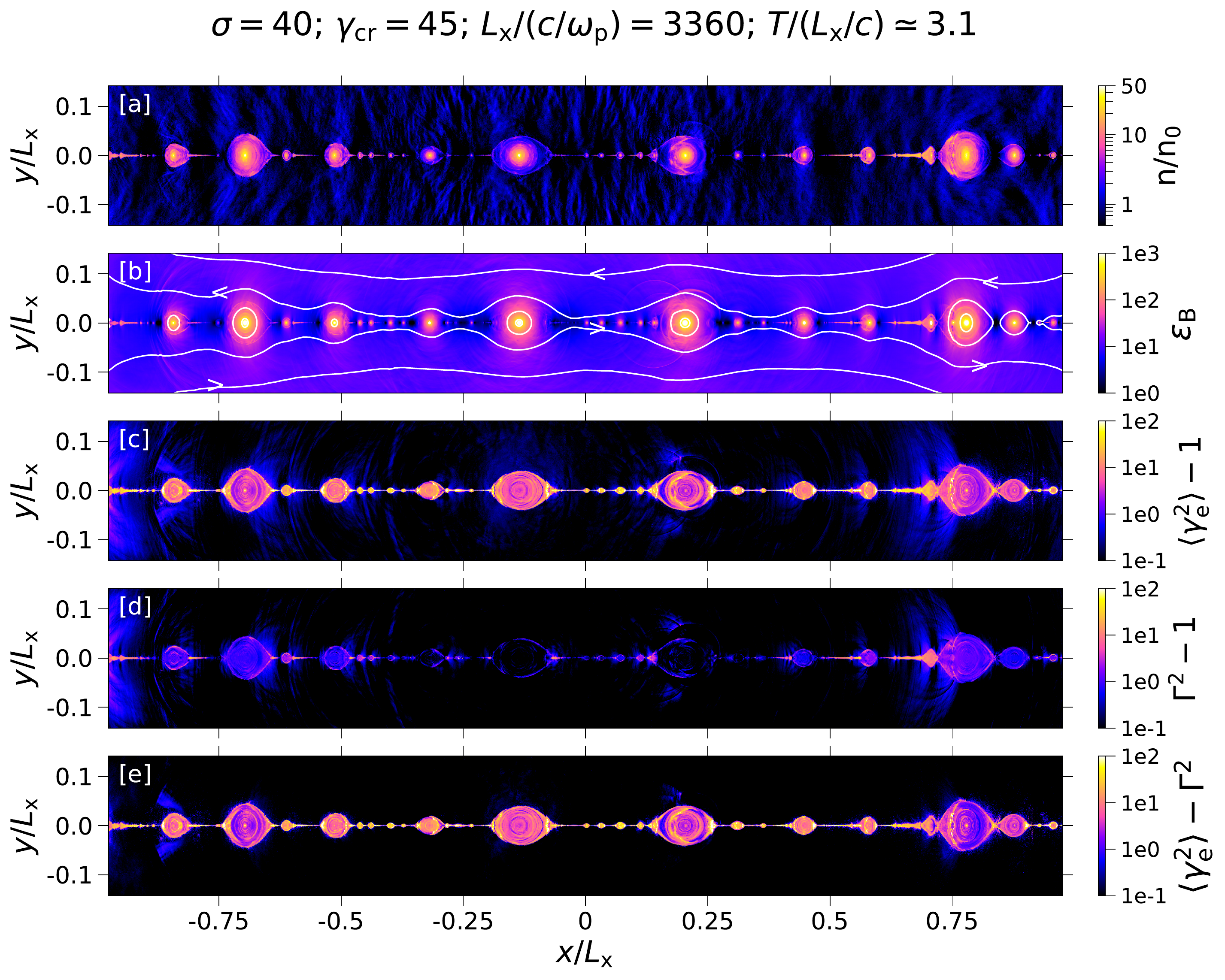}
        \caption*{(ii) Moderate IC cooling}
    \end{subfigure}
    \caption{Snapshot of the reconnection layer at time $T\simeq 3.1 L_{\rm x}/c\simeq 10621~\omega_{\rm p}^{-1}$ for our fiducial model ($\sigma=40$ and $L_{\rm x}=3360c/\omega_{\rm p}$) with strong cooling ($\gcr=16$, left) and moderate cooling ($\gcr=45$, right). We only show the region $|y|\leq 0.15\,L_{\rm x}$ where reconnection occurs. [a] Particle number density $n$ in units of the initialized (upstream) number density $n_{\rm 0}$. [b] Magnetic energy density $B^{2}/8\pi$ normalized to the upstream plasma rest mass density, $\varepsilon_{\rm B}=B^2/8\pi n_{\rm 0}m_{\rm e}c^2$. The over-plotted white contours are magnetic field lines. [c] Local average $\langle\gamma_{\rm e}^2\beta_{\rm e}^2\rangle = \langle\gamma_{\rm e}^2\rangle - 1$, which is proportional to the IC power per particle (the local average is calculated using the patches of $5\times5$ cells). [d] $\Gamma^2-1$, where $\Gamma$ is the bulk Lorentz factor defined in the text. [e] $\langle\gamma_{\rm e}^2\rangle - \Gamma^2$, which represents internal particle motions. The quantities shown in panels [a], [c], [d] and [e] are computed using both electrons and positrons.
}
\label{fig:2D_rec_layer}
\end{figure*}

%%%%%%
\subsection{Structure of the reconnection layer} \label{subsec:general}

In Fig.~\ref{fig:2D_rec_layer}, we present the 2D structure of the reconnection layer in the strongly magnetized model with $\sigma=40$ for two different levels of IC cooling: strong cooling with $\gcr=16$ on the left, and moderate cooling with $\gcr=45$ on the right. Regardless of the cooling strength, the reconnection layer fragments into a hierarchical chain of coherent structures (plasmoids) separated by X-points, as a result of the tearing instability \citep{tajima_shibata_97, loureiro_07, bhattacharjee_09, uzdensky_10}.

Plasmoid mergers lead to the formation of secondary reconnection layers transverse to the main layer. An example of such transverse layers can be seen at $x/L_{\rm x}\sim0.35$ in the left panel of Fig.~\ref{fig:2D_rec_layer}. Overall, the formation of the plasmoid chain proceeds independently of the degree of IC cooling. We have also verified that in our $\sigma=40$ runs, the reconnection rate and the distribution of plasmoid sizes (see Appendix \ref{appendix:plasprop}) are not significantly affected by cooling, similar to the lower magnetization case ($\sigma=10$) discussed by  \citetalias{Sironi_20}.

In the case of moderate cooling (right panel [a] in Fig.~\ref{fig:2D_rec_layer}), particles are nearly symmetrically distributed in plasmoids (i.e., front and back sides are equally populated). On the other hand, under strong cooling, the plasma density distribution inside plasmoids becomes strongly non-uniform (left panel [a]). A near-vacuum region develops at the plasmoid ``head''---along the direction of motion---leaving a ``tail'' of piled-up particles on the back side. This effect is caused by Compton drag on the particles. Since the particles can freely slide along magnetic field lines, Compton drag pushes them toward the back of the moving plasmoid. The cavities are most pronounced inside plasmoids that are farther from the center, because they typically move faster and have lived longer, allowing more time for the drag.

The 2D structure of magnetic energy density ($\epsilon_{\rm B}=B^2/8\pi n_{\rm 0}m_{\rm e}c^2$; panels [b] of Fig.~\ref{fig:2D_rec_layer}) is similar in models with different levels of cooling. In particular, the boundaries of plasmoids, defined as regions of large $\epsilon_{\rm B}$, have similar shapes, and the peak value of $\epsilon_{\rm B}$ inside the plasmoids is also nearly the same. This peak occurs in the cores (O-points) of plasmoids, and their measured values for the largest plasmoids are $\epsilon_{\rm B}\sim 7\times 10^3$. Note that the normalized magnetic energy density in the inflow region is $\epsilon_{\rm B}\sim \sigma/2\sim20$. Hence, the magnetic field in the plasmoid cores has been compressed by a factor of $B/B_{\rm 0}\sim18$ relative to the upstream field. The similarity of plasmoid chains in all cooling models confirms that the chain structure is controlled by magnetic stresses, with small effects of plasma pressure. In short, plasmoids in relativistic reconnection are fundamentally magnetic structures.

In panels [c], we plot the local average $\langle\gamma_{\rm e}^2\beta_{\rm e}^2\rangle=\langle\gamma_{\rm e}^2\rangle-1$, which is proportional to the IC power per particle. In the strongly cooled case, the plasma has lost most of the energy received from reconnection, so $\langle\gamma_{\rm e}^2\beta_{\rm e}^2\rangle$ is strongly reduced. For both strong and moderate cooling, we find that $ \langle\gamma_{\rm e}^2\beta_{\rm e}^2\rangle$ is highest in the thin regions of the reconnection layer (the sites of X-points and fast outflows emanating from them) and the heads and tails of the plasmoids. Note that significant dissipation occurs at the plasmoid boundaries, because of (\textit{i}) mergers and (\textit{ii}) the release of kinetic energy via the slowdown of outflows launched from a neighboring X-point and entering the plasmoid.

In panels [d], we isolate the contribution due to bulk motions alone. For every cell, we calculate the average particle velocity $\bmath{\beta}=\langle\bmath{\beta}_{\rm e}\rangle$ including all electrons and positrons in the local patch of $5\times5$ cells. In the frame moving with velocity $\bmath{\beta}$, the plasma stress-energy tensor has a vanishing energy flux \citep[e.g.,][]{rowan_19}, and we define it as the comoving frame of the plasma. The corresponding bulk Lorentz factor is $\Gamma=(1-\beta^2)^{-1/2}$. We find that the bulk motions of large plasmoids weakly depend on the cooling strength. The plasmoids move nearly as rigid bodies, and their speed on average grows with distance from the center of the layer. This growth is somewhat limited by Compton drag in the model with strongest cooling, which results in a slightly slower outflow. Young plasmoids develop high speeds, as they are pulled toward large plasmoids. Fast bulk motions also occur in the tails of plasmoids that accrete the reconnected plasma from nearby X-points (see e.g. the plasmoid at $x/L_{\rm x}\sim-0.7$ in the right panel). Besides the motions along the layer, we observe high bulk speeds in the secondary reconnection layers generated by plasmoid mergers (e.g., at $x/L_{\rm x}\sim0.35$ in the left panel).

In the case of strong cooling (left column), the similarity between panels [c] and [d] suggests that most of the IC power comes from bulk motions. The effect of 
Internal motions of individual particles may be quantified by the difference $\langle\gamma_{\rm e}^2\rangle-\Gamma^2$ (panel [e]). In the case of moderate cooling, internal motions dominate over bulk motions almost everywhere in the reconnection layer. In the model with strong cooling, internal motions are suppressed
inside plasmoids down to a non-relativistic temperature. This is especially true for large plasmoids, which are long-lived, so their particles have plenty of time to radiate away their energy.\footnote{The effect of IC cooling on plasmoids as a function of their size in investigated in Appendix \ref{appendix:plasprop}.} High-temperature internal motions are predominant in the thin regions of the main reconnection layer (and secondary layers generated by plasmoid mergers) where particles have been recently energized and are yet to cool down.\footnote{Significant internal motions are also observed near the core of merging plasmoids, likely due to compression (e.g., see the bright regions in panel [e] of the left column at $x/L_{\rm x}\sim0.3$ and 0.9).} As shown by \citetalias{Sironi_20}, particles accelerated at X-points or ``picked up'' by the fast reconnection outflows are energized on a timescale shorter than the IC cooling time.

%%%%%%%%%%%%%%%%%%
\subsection{Bulk and internal motions} \label{subsec:velocities}

We now explore in more detail how the bulk and internal motions depend on the flow magnetization and the cooling level. We exclude the cold, slow inflow and examine motions in the ``reconnection region,'' which is defined using the following density-mixing criterion \citep[e.g.,][]{rowan_17, ball_18}: the region that contains a mixture of particles coming from above and below the reconnection midplane $y=0$, with both populations contributing at least 1\% to the mixture. All results presented below refer to particles and cells from this region unless otherwise specified.

\begin{figure*}
    \centering 
    \begin{subfigure}[t]{0.5\textwidth}
        \centering
      \includegraphics[width=0.963\textwidth]{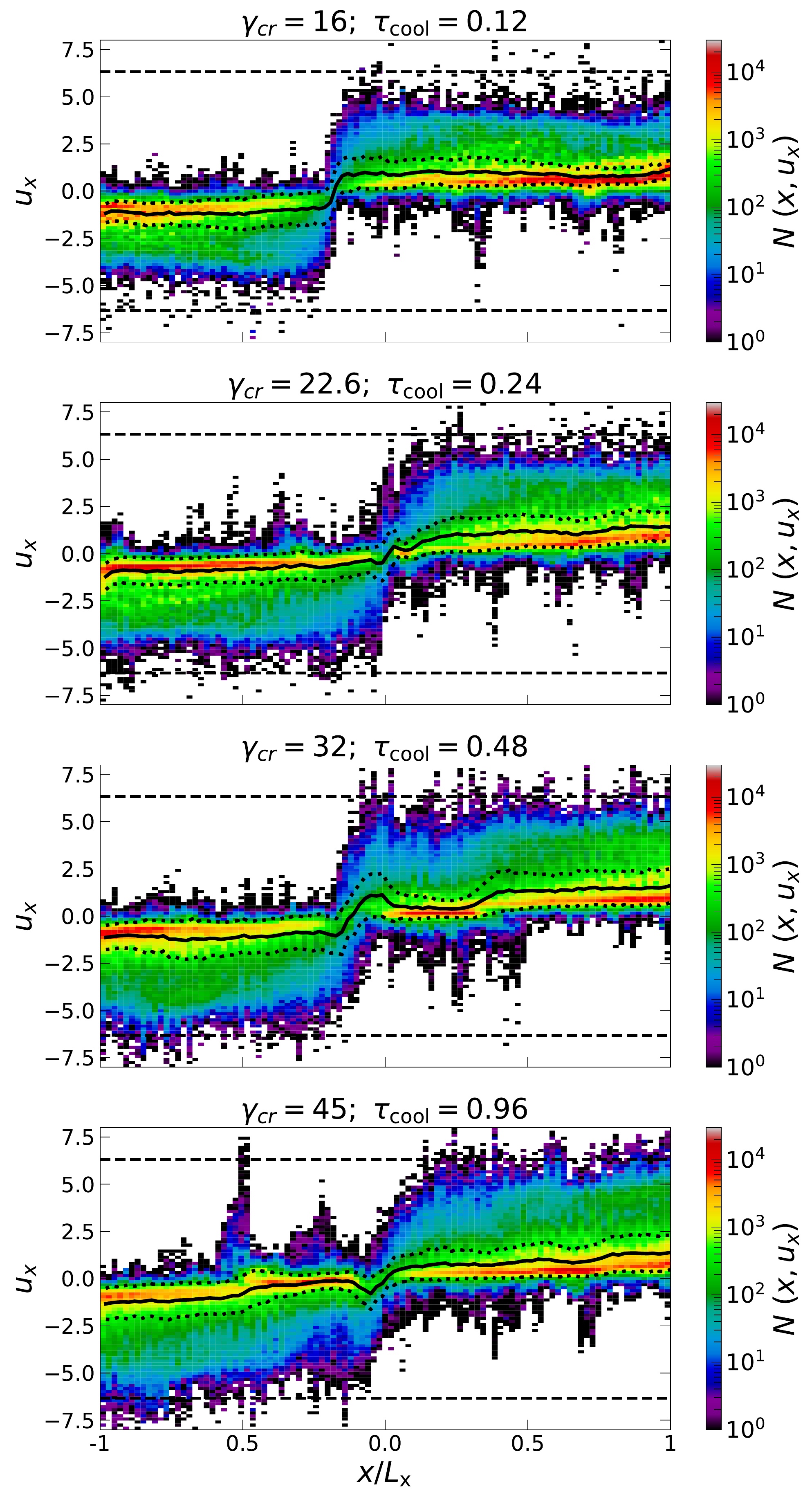}
    \end{subfigure}%
    ~ 
    ~
    \begin{subfigure}[t]{0.5\textwidth}
        \centering
        \includegraphics[width=1.0\textwidth]{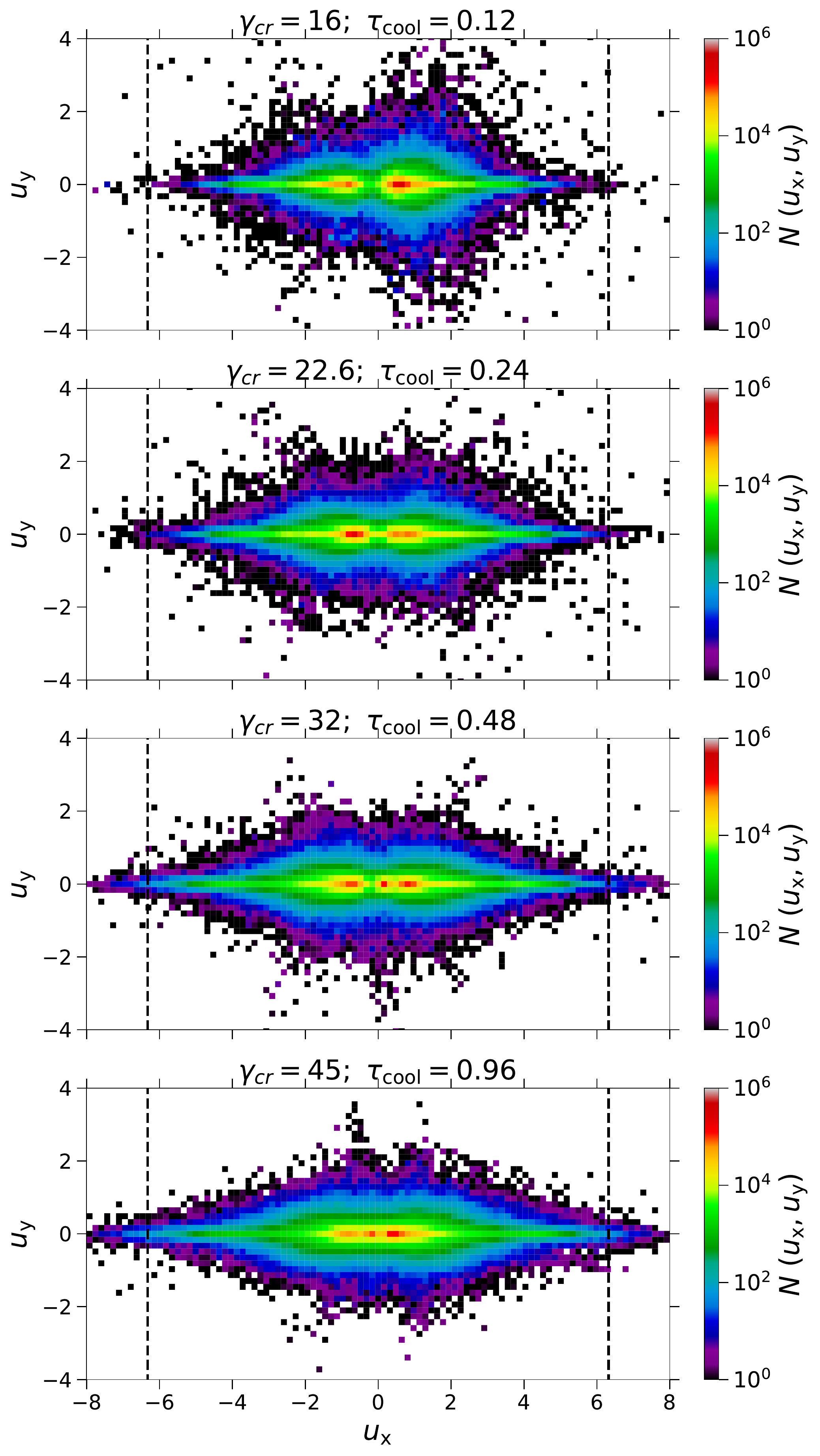}
    \end{subfigure}
    \caption{
\textit{Left:} bulk motions along the reconnection layer, viewed in the phase space $x-\ux$. Color represents particle density in the phase space. The measurements were made for our fiducial model with $\sigma=40$, $L_{\rm x}/(c/\omega_{\rm p})=3360$, and different panels correspond to different levels of cooling: $\gamma_{\rm cr}=16,22.6,32,45$ (strongest to weakest from top to bottom). The solid black curve in each panel shows the mean value of $\ux$ as a function of $x$, and the black dotted curves show the standard deviation around the mean. \textit{Right:} bulk motions along $x$ and $y$, viewed in the momentum space $\ux-\uy$, for the same simulations as in the left panels. Color represents the particle density in the $\ux-\uy$ space. In all (left and right) panels, the black dashed lines indicate the Alf\`enic limit $\sqrt{\sigma}$. The distributions were time-averaged during the time interval $2\lesssim T/(L_{\rm x}/c)\lesssim5$, when the reconnection layer was in a quasi-steady state.
}
\label{fig:phase-space}
\end{figure*}

First, we calculate the bulk 4-velocities parallel ($u_{\rm x}=\Gamma\beta_{\rm x}$) and orthogonal ($u_{\rm y}=\Gamma\beta_{\rm y}$) to the reconnection layer. 
Fig.~\ref{fig:phase-space} shows the distributions of $u_{\rm x}$ and $u_{\rm y}$ measured across different cells. To ensure sufficient statistics, we only use cells containing $\ge4$ particles, and the distributions are averaged over time interval $2\lesssim T/(L_{\rm x}/c)\lesssim 5$, when reconnection proceeds in a quasi-steady state. Fig.~\ref{fig:phase-space} presents the resulting distributions in the $x-u_{\rm x}$ space (left) and in the $u_{\rm x}-u_{\rm y}$ space (right), for a sequence of models with different cooling levels.

The observed behavior in our fiducial model with $\sigma=40$ is similar to the model with $\sigma=10$ described in \citetalias{Sironi_20}. Bulk motions are primarily oriented along the layer, i.e., they are dominated by $\ux$. The mean value of $\ux(x)$ is nearly independent of the cooling level. It starts from zero near the center of the layer and approaches a constant value $|\ux|\sim 1$ at $|x|/L_{\rm x}\gtrsim 0.25$. Most of the reconnected plasma moves with trans-relativistic $|\ux|\sim 1$. Only a small fraction moves with ultra-relativistic bulk speeds, approaching the Alfv\'enic limit \citep{lyubarsky_05}: $|u_{\rm x}|\approx\sqrt{\sigma}\approx6.3$ (only small plasmoids can reach  $|u_{\rm x}|\approx\sqrt{\sigma}$, as discussed in Appendix~\ref{appendix:plasprop}). We note that this fast fraction is reduced for (i) stronger cooling, due to Compton drag, and (ii) larger magnetization $\sigma=40$, compared with the $\sigma=10$ model investigated by \citetalias{Sironi_20}; a similar effect was discussed by \citet{sironi_16} for uncooled simulations.

A fraction of the reconnected plasma is flowing towards the center of the layer (i.e., opposite to the mean motion). This is seen in \fig{phase-space} [left] as localized spikes with $\ux<0$ at $x>0$ (and $\ux>0$ at $x<0$). These motions are caused by large plasmoids attracting and accreting small young plasmoids ahead of them. As a result, a small fraction of plasma motions opposite to the mean outflow motion.

\begin{figure} 
\includegraphics[width=8cm]{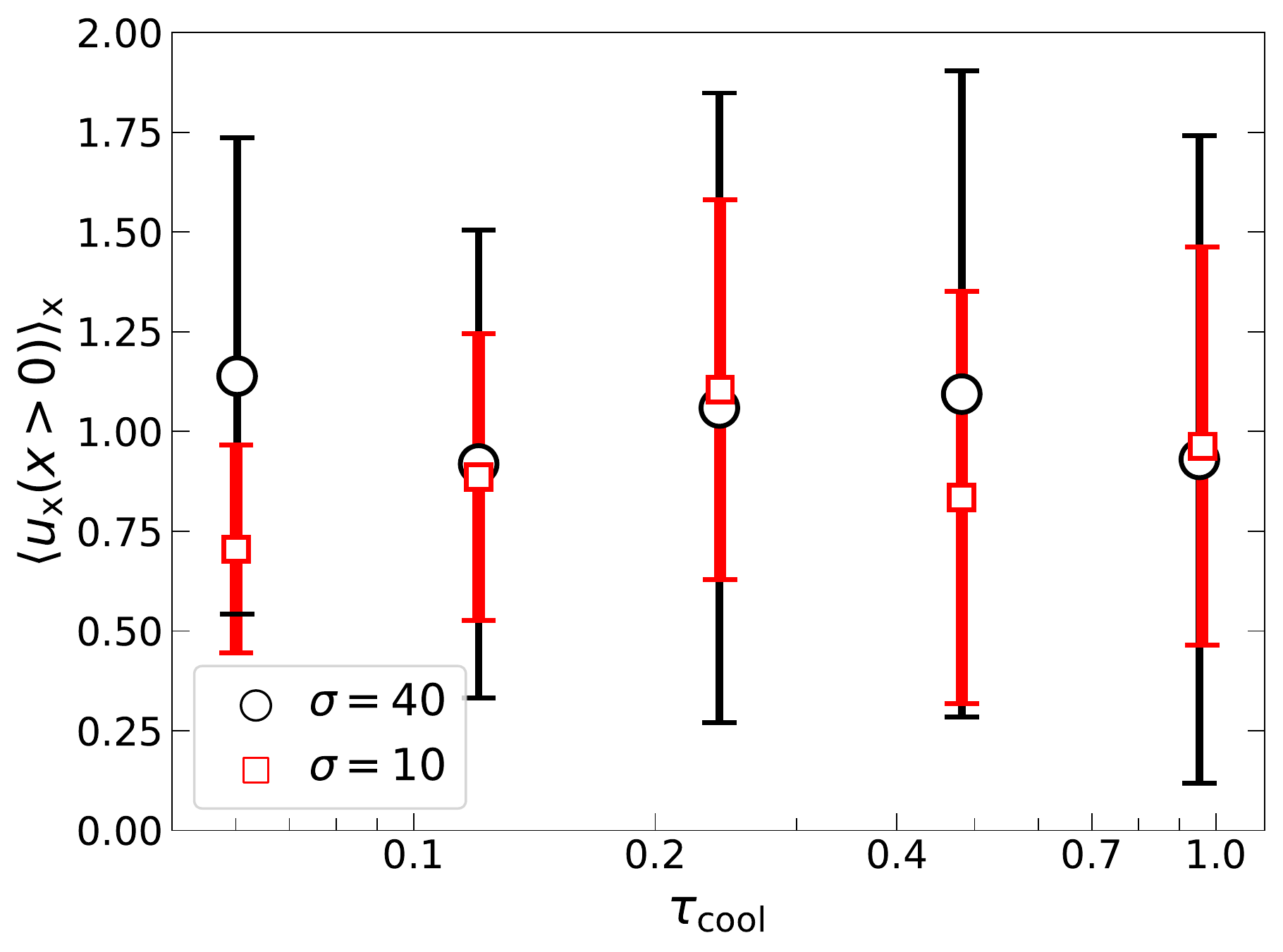}
\caption{
Statistics of bulk motions along the reconnection layer in the models with various $\tau_{\rm cool}$, for $\sigma=10$ (red) and $\sigma=40$ (black). Red squares and black circles indicate the mean (density-weighted) bulk 4-velocity $\ux$ for $\sigma=10$ and 40 respectively, and the error bars indicate its standard deviation. The measurements were performed along one half of the reconnection layer ($x>0$) and averaged during the time interval $2 \lesssim T/(L_{\rm x}/c) \lesssim 5$.
} 
\label{fig:std_ux}
\end{figure}

The stochastic character of plasmoid motions plays an important role for Comptonization calculations (see \S\ref{sec:radiative_transfer}) and it is further quantified in Fig.~\ref{fig:std_ux}. We use half of the reconnection region ($x>0$), and compute the global mean and standard deviation of the bulk 4-velocity $\ux$ for $\sigma=10$ and $40$ and for a variety of cooling levels, $0.06\lesssim\gcool\lesssim1.0$. For each $\sigma$ and $\tau_{\rm cool}$, we show the result for the largest available simulation (which also has the highest $\gcr$, see \eq{varrho}). We verified that the results remain nearly the same when varying $L_{\rm x}$ and $\gcr$ for given $\sigma$ and $\gcool$ (see Appendix~\ref{appendix:table}).

One can see from Fig.~\ref{fig:std_ux} that the global mean $\langle\ux\rangle\sim 1$ is similar in all models, and the standard deviation has a significant systematic dependence on both $\gcool$ and $\sigma$: it increases for larger $\gcool$ (i.e., weaker cooling) and for higher $\sigma$. Thus, bulk motions are more ordered for lower magnetizations and stronger cooling. The dispersion in bulk motions for $\sigma=40$ is roughly twice that for $\sigma=10$, at a given $\gcool$.

Bulk motions in the $y$ direction (orthogonal to the main reconnection layer) are quantified by the distribution in the $\ux-\uy$ space (\fig{phase-space} [right]). The $\uy$ component is mostly produced by the secondary transverse reconnection layers created during plasmoid mergers. Most mergers occur between plasmoids with moderate $x$-velocities, which explains why the largest $\uy$ are observed at moderate $\ux$.  However, overall the $y$-motions in the reconnection layer are weaker than the $x$-motions, and $\uy$ never reaches the Alfv\'en limit. This may be explained by the fact that the transverse reconnection layers are shorter than the main layer (their extent is limited by the width of the largest plasmoids, $\sim 0.1\,L_{\rm x}$), and the reconnected plasma may fail to reach the expected Alfv\'en limit. In addition, the secondary reconnection occurs in already heated plasma, which tends to reduce the effective magnetization and the corresponding Alfv\'en speed in the layers between merging plasmoids. However, strong cooling increases the local magnetization, and so models with lower $\gcool$ display faster $y$-motions. Furthermore, strong Compton drag creates density cavities (see \S\ref{subsec:general}), boosting the local magnetization parameter. Overall, the reduced temperature and density result in faster reconnection outflows. 

%%%%%%%%%%%
%\subsubsection{Bulk vs. internal energy} \label{subsubsec:energies}

Next, we examine how the energy in bulk motions compares with the internal energy of the plasma (heat). In each cell, the mean bulk energy per particle (in units of the rest mass energy) is simply $\Gamma-1$. The mean internal energy per particle is obtained as described in Appendix \ref{appendix:internal} \citep[e.g.,][]{rowan_17}. We Lorentz boost the particle energy density from the simulation frame to the plasma rest frame, which moves at velocity $\boldsymbol{\beta}$. The mean internal energy per particle $\Eint $ (normalized to the rest mass energy) is then calculated assuming that the pressure tensor is isotropic in the plasma rest frame. Using the values obtained in each cell, we find the global mean values $\langle\Gamma-1\rangle$ and $\langle\Eint \rangle$, first by (density-weighted) averaging over the reconnection region, and then by time-averaging over the interval $2\lesssim T/(L_{\rm x}/c)\lesssim5$.

\begin{figure} 
\includegraphics[width=8cm]{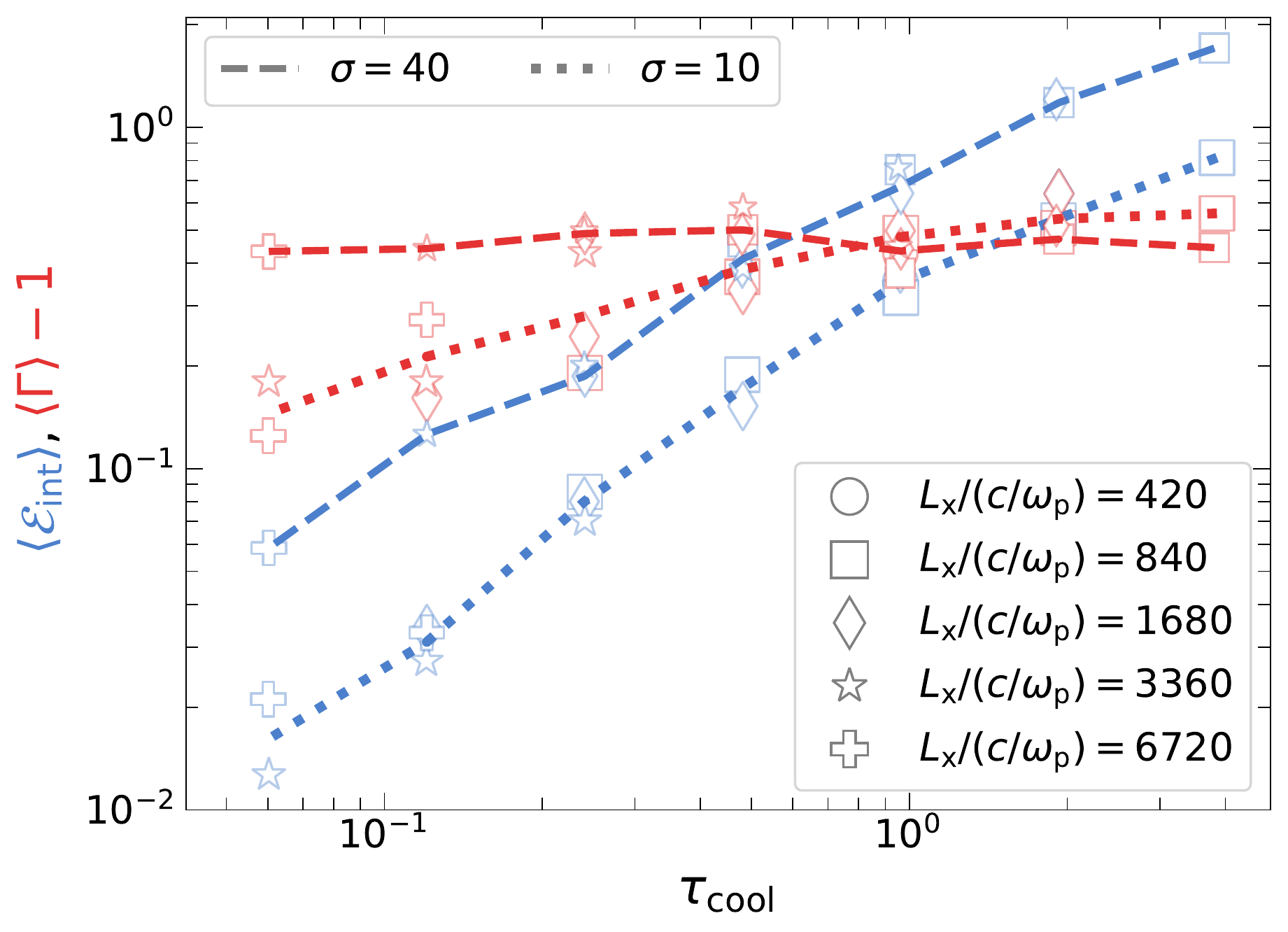}
\caption{
Average particle energy (in units of $m_{\rm e}c^2=511$~keV) resolved into two components: internal (blue) and bulk (red), plotted vs. $\gcool$. Dashed curves refer to $\sigma=40$, and dotted to $\sigma=10$. For given $\sigma$ and $\gcool$, different symbols correspond to different sizes of the simulation domain (see legend), which equivalently correspond to different $\gcr$. Where more than one simulation is available with the same $\sigma$ and $\gcool$ (Table~\ref{tab:setup}), the curve passes through the mean value. The measurements were performed using density-weighted averaging over the reconnection region and averaging over time interval $2 \lesssim T/(L_{\rm x}/c) \lesssim 5$.}
\label{fig:energies}
\end{figure}

\fig{energies} shows how the measured $\langle\Gamma-1\rangle$ and $\langle\Eint \rangle$ depend on $\gcool$.\footnote{\fig{energies} also demonstrates that variations in $\gcr$ and $L_{\rm x}$ do not appreciably change the results as long as they give the same $\gcool$. This fact is not surprising, as variation in $\gcr\gg\sigma/4$ directly affects only very energetic particles, with Lorentz factors far beyond the mean post-reconnection value $\sim \sigma/4$.} 
The similarity of bulk motions in all models of $\sigma\gg1$ and IC cooling results in a nearly universal $\langle\Gamma-1\rangle\sim 0.4$. The only exception is the decrease observed in the strongest cooling cases of $\sigma=10$, down to $\langle\Gamma-1\rangle\sim 0.1$ for $\gcool=0.06$. This decrease is the result of Compton drag. Since the ratio of drag time to advection time is $t_{\rm drag}/t_{\rm adv}\sim \sigma\,\gcool$ (see Eqs.~(\ref{eq:t_drag}) and \S\ref{subsec:time_scales}), we expect Compton drag to affect the average bulk motions when $\gcool\lesssim \sigma^{-1}$.

As one can see from \fig{energies}, the ``hot regime'' with  $\langle\Eint \rangle>\langle\Gamma-1\rangle$ occurs for large $\gcool$, namely $\gcool\gtrsim 1$ for $\sigma=10$ and $\gcool\gtrsim 0.5$ for $\sigma=40$. In the limit of negligible cooling losses, the mean internal energy per particle $\langle\Eint \rangle$ should approach the mean Lorentz factor $\sim \sigma/4$ of particles heated by reconnection. We speculate that this asymptotic limit should be approached when $\gcool\gtrsim \sigma/4$, i.e., when the majority of particles heated/accelerated by reconnection do not appreciably cool before advecting out of the system.\footnote{We remind that for $\gcool>1$, the cooling parameter $\gcool$ equals the Lorentz factor of particles that cool in a dynamical (advection) time.}

%%%%
\subsection{Particle energy spectra} \label{subsec:spectra}

In this section, we investigate how the particle energy spectrum varies with $\sigma$ and cooling strength. Note that the cooling effects depend on the cooling rate (parameterized by $\gcr$) and the characteristic time $L_{\rm x}/c$ available for cooling before the particles are advected out of the layer. Therefore, we examine the dependence on both $\gcr$ and $L_{\rm x}$.

For each particle, we identify the energy contribution from the local bulk motion. Then, we can compare the spectrum of bulk motions with the total energy spectrum and see where bulk motions dominate. The spectra are extracted from the reconnection region and averaged in time over the interval $2\lesssim T/(L_{\rm x}/c) \lesssim 5$, when the layer has achieved a quasi-steady state. They are normalized by $L_{\rm x}^2$ to allow a fair comparison of simulations with different $L_{\rm x}$.

\begin{figure*} 
\includegraphics[width=16cm]{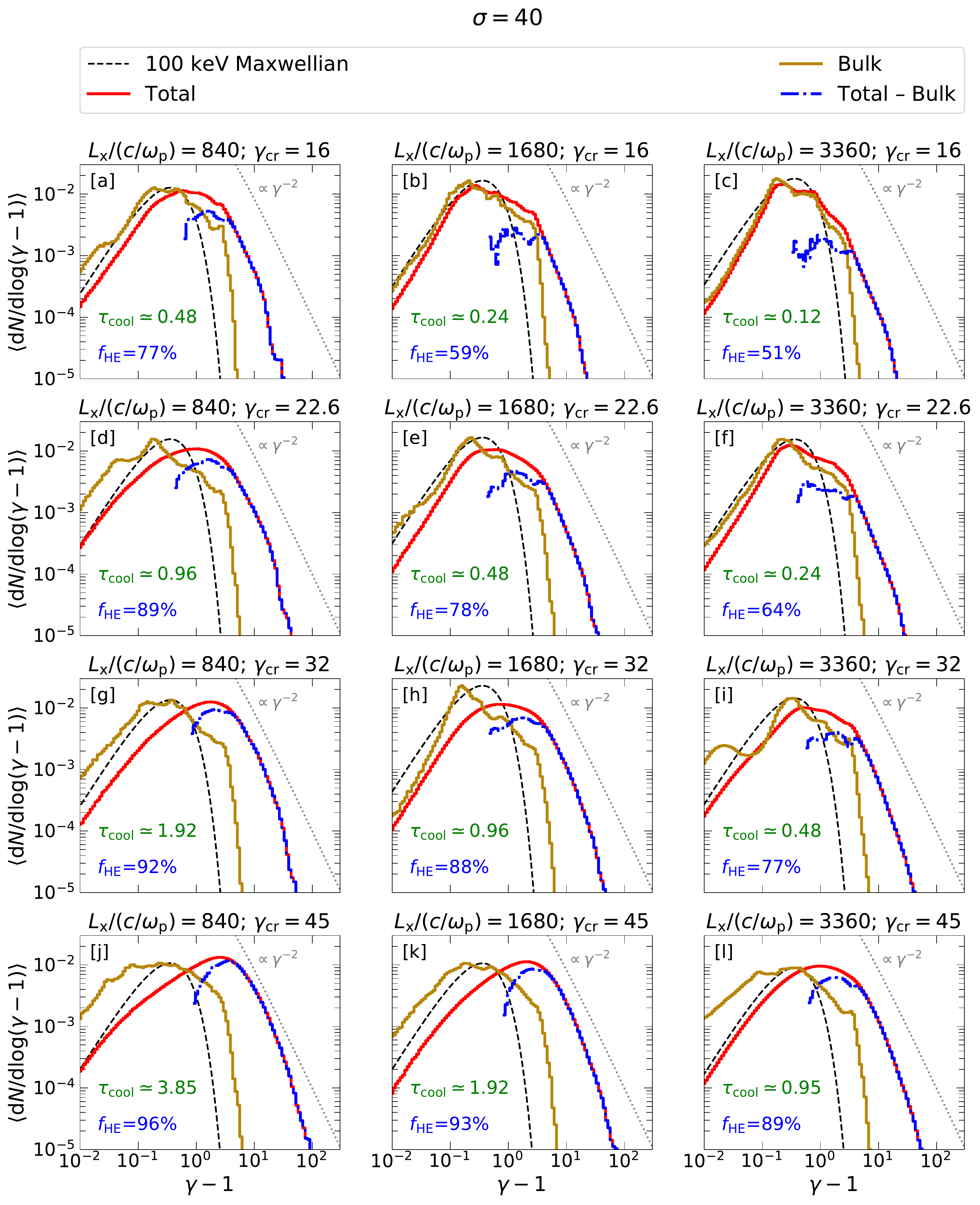}
\caption{Particle energy spectra extracted from the reconnection region and averaged over time $2\lesssim T/(L_{\rm x}/c)\lesssim 5$. Twelve simulations are shown with
magnetization $\sigma=40$ and different values of $\gcr$ and domain size $L_{\rm x}$. The background radiation density decreases from top to bottom ($\gcr$ increases from 16 to 45) while the domain size increases from left to right (from $L_{\rm x}=840\,c/\omega_{\rm p}$ to $L_{\rm x}=3360\,c/\omega_{\rm p}$). The corresponding value of $\gcool$ is indicated in each panel. The spectra are normalized by $L_{\rm x}^2$ for proper comparison of models with different $L_{\rm x}$. Each panel shows 
the total energy spectrum (red; $\gamma=\gammae$), the bulk motion spectrum (golden-brown; $\gamma=\Gamma$), and their difference (dash-dotted blue), i.e. the part not accounted for by bulk motions; the fractional IC power contributed by this component is denoted in each panel as $f_{\rm HE}$. For comparison, we also plot a Maxwellian distribution with temperature of 100 keV (dashed black), normalized so that its peak matches the peak of the bulk energy spectrum. The dotted grey lines indicate the $-2$ slope that corresponds to equal IC power per decade in Lorentz factor.
}
\label{fig:spectra_domain}
\end{figure*}

\begin{figure*} 
\includegraphics[width=16cm]{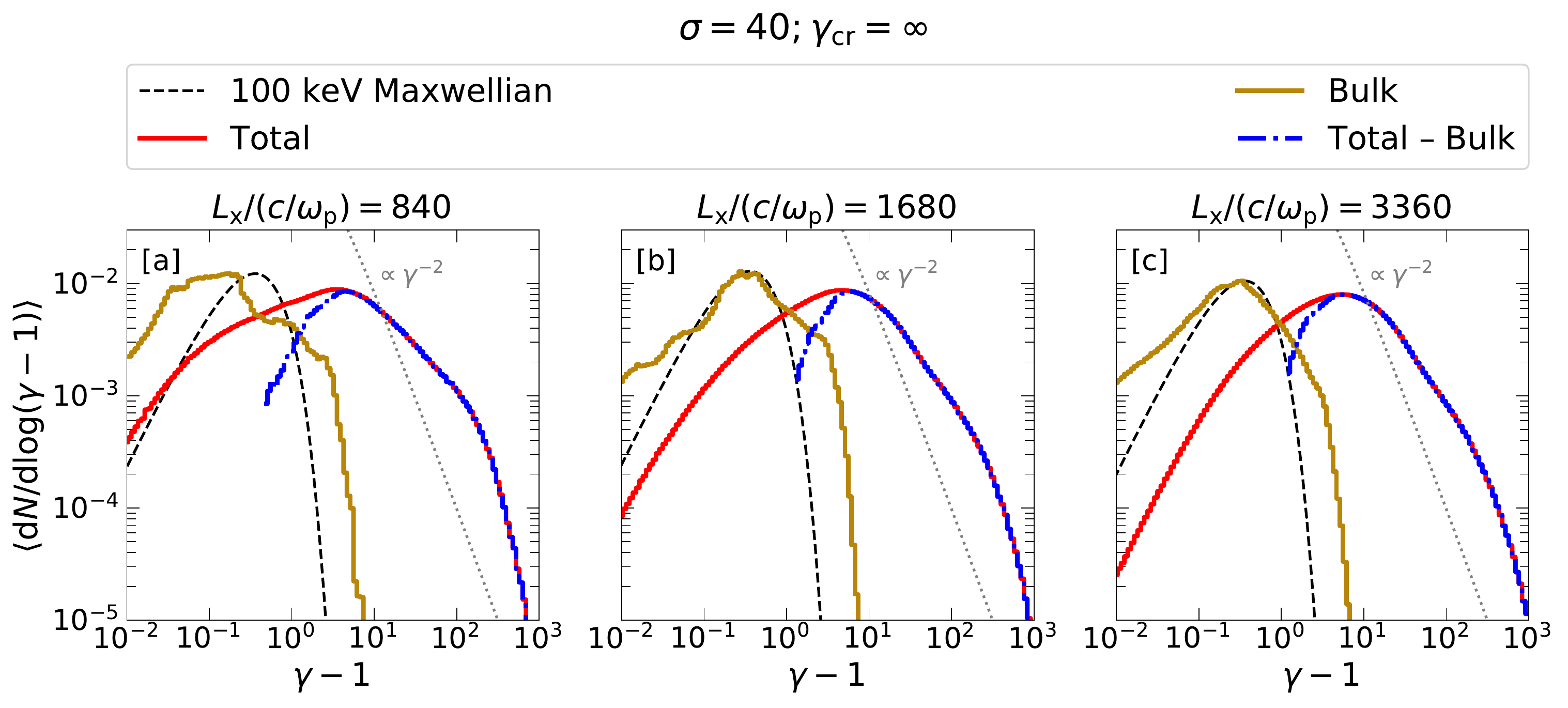}
\caption{Same as in each row of Fig.~\ref{fig:spectra_domain}, but for the  simulations without cooling ($\gcr=\infty$).}
\label{fig:spectra_domain_nocool}
\end{figure*}

\fig{spectra_domain} presents the results for $\sigma=40$. Each row corresponds to a chosen value of $\gcr$ (increasing from top to bottom; so, the cooling rate
decreases from top to bottom), and each column has a chosen domain size (increasing from left to right). The parameter $\gcool\propto \gcr^2/L_{\rm x}$ varies diagonally (bottom-left to upper-right) in this $4\times 3$ set of models. For completeness, we present in \fig{spectra_domain_nocool} the corresponding simulations without cooling, i.e., $\gcr=\infty$. 

In the absence of IC cooling (Fig.~\ref{fig:spectra_domain_nocool}), the total particle spectrum peaks at $\gamma_{\rm e}-1\sim10\sim\sigma/4$ regardless of the domain size, and the mean Lorentz factor of particles heated by reconnection is determined by energy conservation, $\bar{\gamma}_e\approx \sigma/4$. With increasing cooling, the spectral peak progressively shifts to lower values (Fig.~\ref{fig:spectra_domain}). Its position and shape are primarily controlled by $\gcool$. In fact, the total particle spectra for simulations with the same $\gcool$ are remarkably similar. At $\gcool\ll 1$, the strong cooling results in a non-relativistic temperature (see also \fig{energies}), and the total energy spectrum converges towards the bulk energy spectrum, i.e., the particle energies are dominated by the bulk motions. This is especially clear in the top right panel, which has the lowest $\gcool\simeq0.12$.

While the total energy spectrum changes drastically for different levels of  cooling, the bulk energy spectrum is barely affected. In the absence of IC cooling ($\gcr=\infty$), the peak of the distribution of bulk motions is typically at  $\Gamma-1\sim0.3$ (Figs. \ref{fig:spectra_domain_nocool}[b,c]).\footnote{The fact that the bulk spectrum peaks at lower energies in the smallest uncooled box ($L_{\rm x}/(c/\omega_{\rm p})=840$ in Fig.~\ref{fig:spectra_domain_nocool}[a]) is due to the stochastic formation of a nearly-stationary ``monster'' plasmoid. In the simulations with cooling, the slow monsters also produce local maxima at low $\Gamma-1\lesssim 0.1$ (Fig.~\ref{fig:spectra_domain}).} Remarkably, the peak of the bulk spectrum is approximately the same ($\Gamma-1\sim 0.3$) regardless of the cooling strength. This behavior of bulk motions is in agreement with \fig{energies}. For all our simulations, the shape of the bulk spectrum resembles a Maxwell--J\"{u}ttner distribution, and its effective temperature is not far from $kT_{\rm e}\sim100$ keV. Thus, the spectrum of bulk motions of a Compton-cooled plasmoid chain mimics a thermal plasma with a mildly relativistic temperature.

At high energies, the total particle spectrum cannot be accounted for by bulk motions alone; instead, it becomes dominated by internal motions (\fig{spectra_domain} and \fign{spectra_domain_nocool}). The high-energy particles are observed in the thin regions of the main reconnection layer and the secondary layers generated by plasmoid mergers (see Fig.~\ref{fig:2D_rec_layer}[e], left). They are present even in the case of strongest cooling ($\gcool\ll 1$), and such particles need to be sustained by a process of non-thermal particle acceleration operating on timescales shorter than the cooling time. Their origin is described by \citetalias{Sironi_20}: the particles are energized either by the non-ideal reconnection electric field at X-points or by being picked up by fast reconnection outflows from the X-points.

\begin{figure}
\includegraphics[width=8cm]{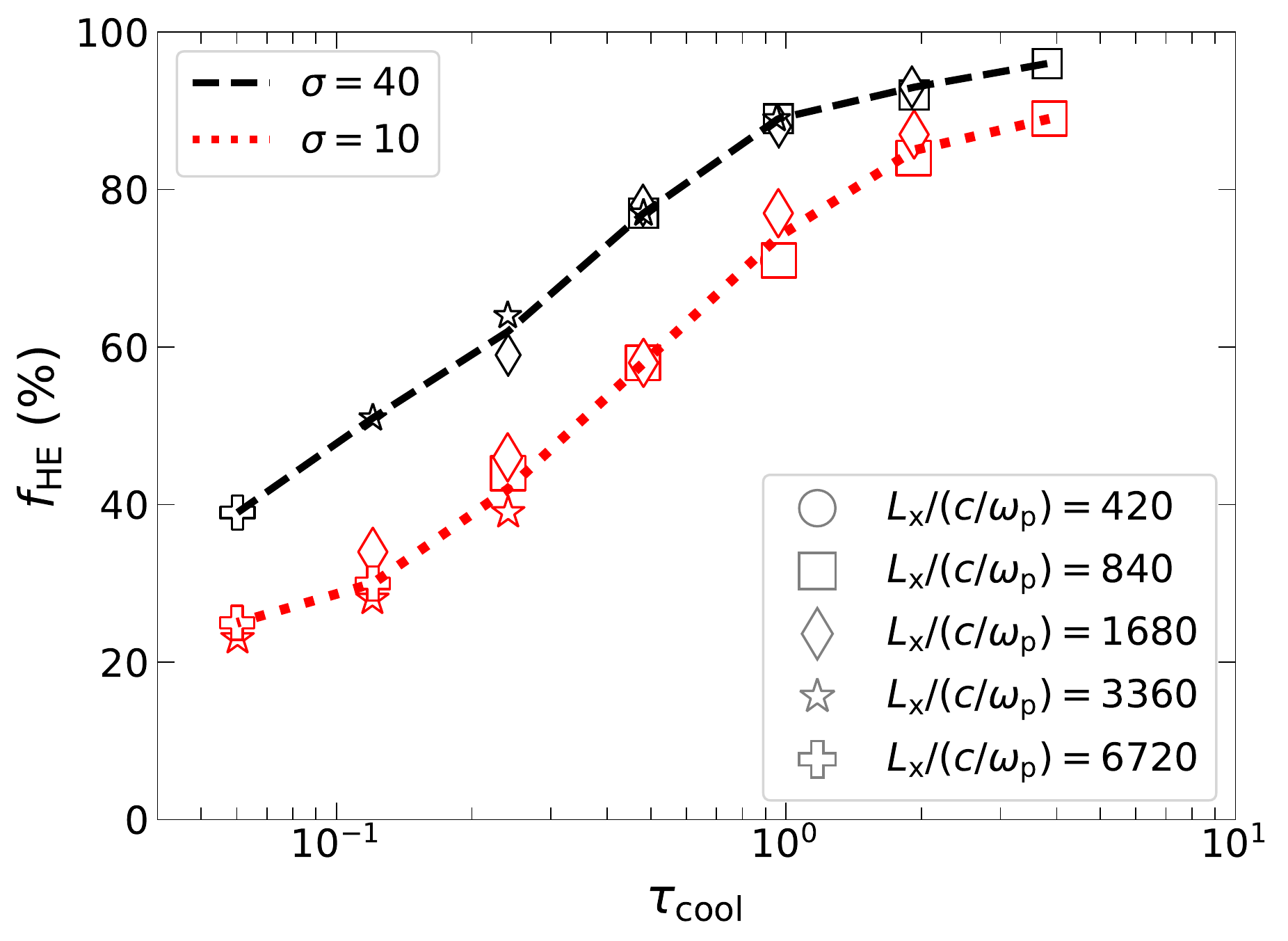}
\caption{
Fraction of the IC power contributed by high-energy particles $f_{\rm HE}$ vs. $\gcool$, measured in the simulations with  $\sigma=10$ (red dotted) and $\sigma=40$ (black dashed). Each data point was calculated by averaging over the reconnection region and then averaging over the time interval $2 \lesssim T/(L_{\rm x}/c) \lesssim 5$. Different symbols represent simulations with different domain sizes $L_{\rm x}$ (see legend) or, equivalently, different $\gcr$. The curves show the average result of all simulations available for given $\sigma$ and $\gcool$.
}
\label{fig:f_HE}
\end{figure}

We define the high-energy component of the particle spectrum as the difference between the total and bulk spectra, as shown in \fig{spectra_domain} and \fign{spectra_domain_nocool}. This component radiates an interesting fraction $\fHE$ of the total IC luminosity from the reconnection layer. The value of $\fHE$
is quoted in each panel of \fig{spectra_domain} and also presented in \fig{f_HE}, where we compare the results for $\sigma=40$ and $\sigma=10$. For both magnetizations, the high-energy component is dominant ($\fHE\gtrsim 90\%$) in the limit of weak cooling, and drops with increasing cooling level. In the simulations with the lowest $\gcool=0.06$, $\fHE\sim25\%$ for $\sigma=10$ and $\fHE\sim40\%$ for $\sigma=40$. \fig{f_HE} also demonstrates that $\fHE$ is controlled by $\gcool\propto \gcr^2/L_{\rm x}$ rather than separately by $\gcr$ and $L_{\rm x}$.

\subsection{Hybrid experiment: cooled electrons and uncooled positrons} \label{subsec:posicool}

\begin{figure*}
    \centering 
    \begin{subfigure}[t]{0.5\textwidth}
        \centering
        \includegraphics[height=2.8in]{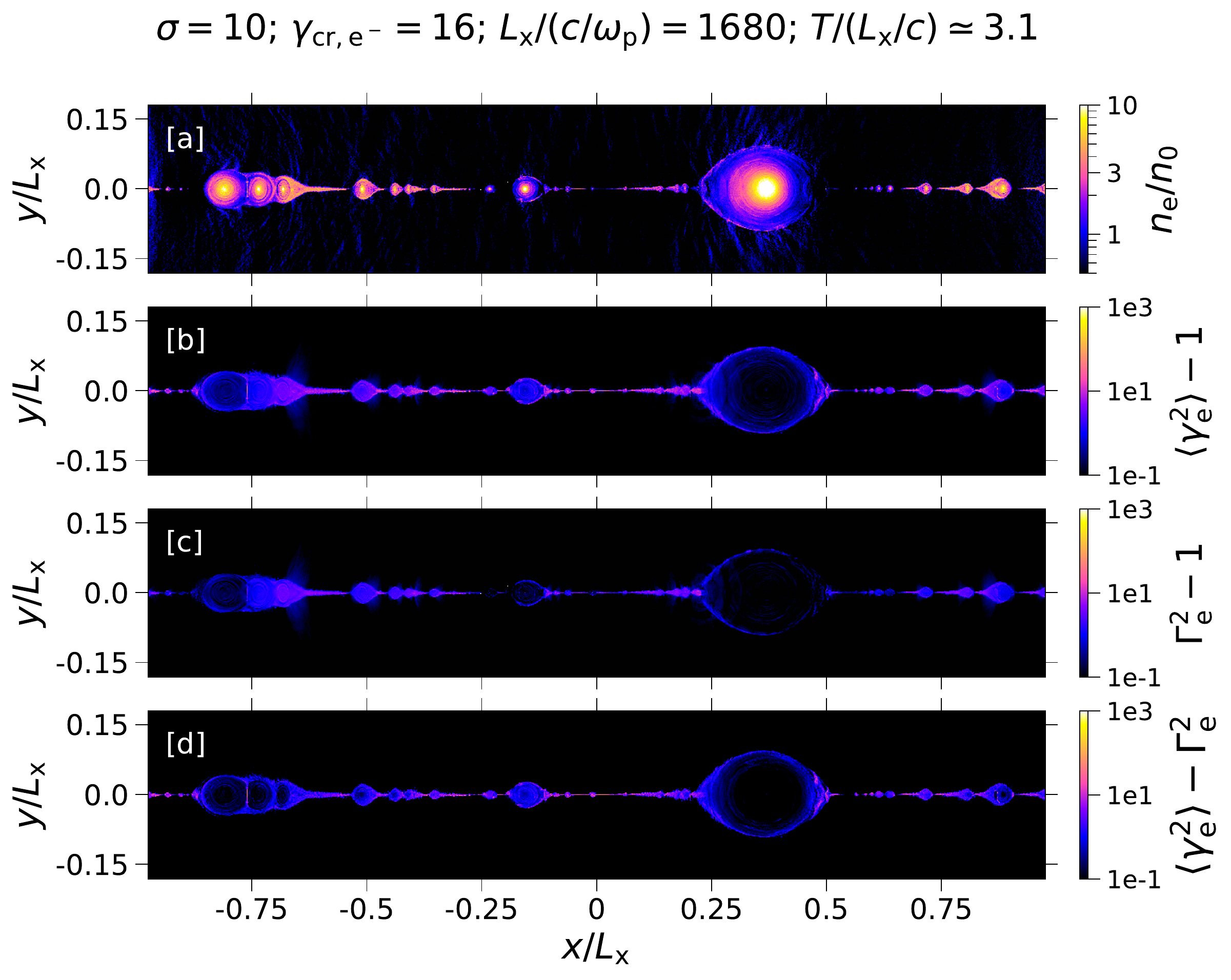}
        \caption*{(i) Electrons}
    \end{subfigure}%
    ~ 
    \begin{subfigure}[t]{0.5\textwidth}
        \centering
        \includegraphics[height=2.8in]{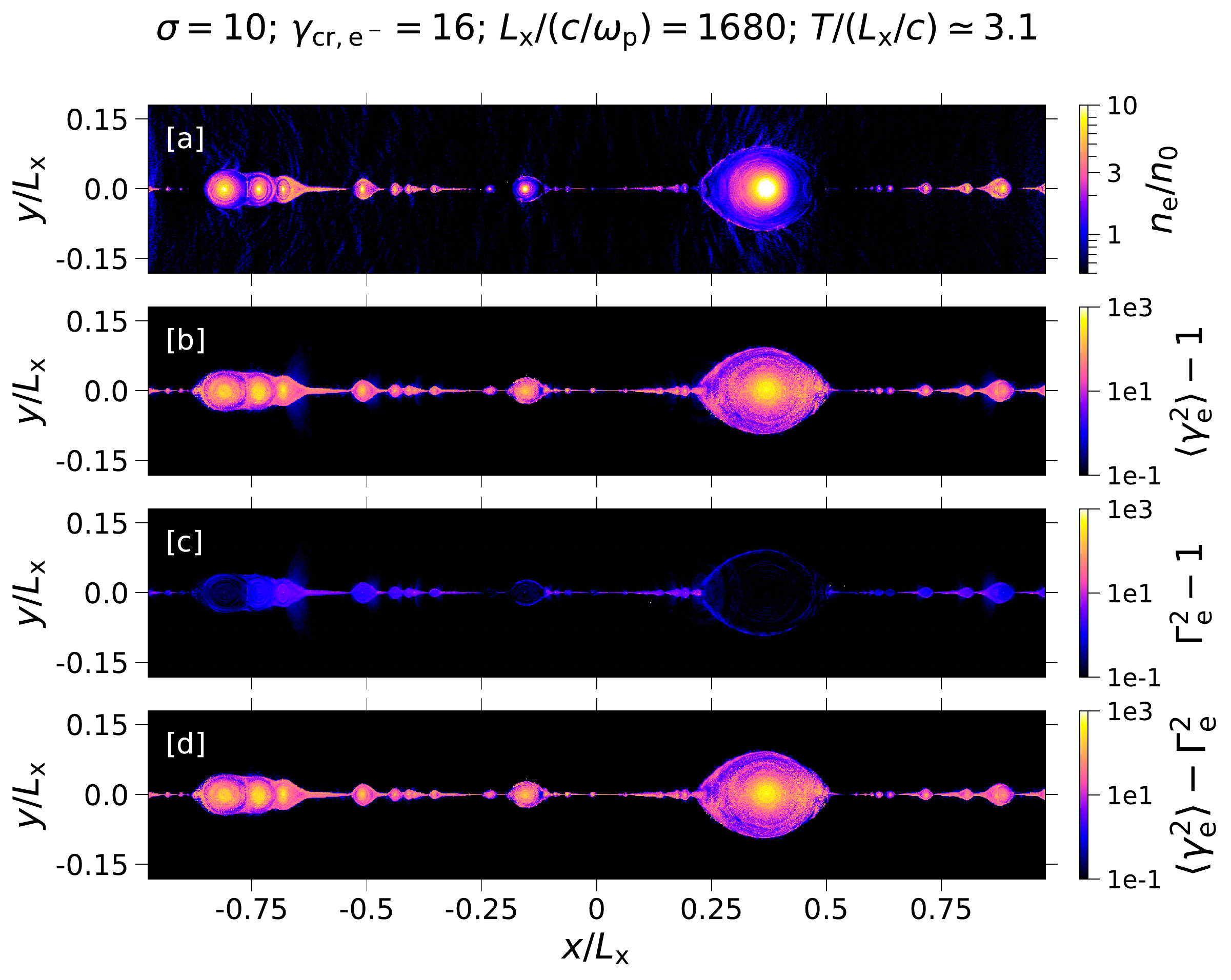}
        \caption*{(ii) Positrons}
    \end{subfigure}
    \caption{2D snapshot of the reconnection layer at time $T\simeq 3.1~L_{\rm x}/c\simeq 10621~\omega_{\rm p}^{-1}$ for electrons (left) and positrons (right), in the hybrid simulation with magnetization $\sigma=10$ and box size $L_{\rm x}/(c/\omega_{\rm p})=1680$. Only electrons are cooled, with a radiation energy density parameterized by $\gamma_{\rm cr, e^-}=16$. [a] Particle number density $n$, in units of the initialized (upstream) number density $n_{\rm 0}$; [b] Local average $\langle\gamma_{\rm e}^2\beta_{\rm e}^2\rangle = \langle\gamma_{\rm e}^2\rangle - 1$, which is proportional to the IC power per particle (the local average is calculated using the patches of $5\times5$ cells). [c] $\Gamma^2-1$, where $\Gamma$ is the bulk Lorentz factor. [d] $\langle\gamma_{\rm e}^2\rangle - \Gamma^2$, which represents internal particle motions.}
\label{fig:2D_rec_layer_posicool_ion}
\end{figure*}

Magnetic reconnection in electron-ion plasmas gives the
electrons the average energy $\sim \sigma_{\rm e}m_{\rm e} c^2$, where $\sigma_{\rm e}=B_{\rm 0}^2/4\pi n_{\rm e}c^2$ is the ``electron magnetization'' \citep{rowan_17,rowan_19,werner_18,petropoulou_19}. Bulk motions are primarily determined by the ``ion magnetization'' $\sigma_{\rm i}=(m_{\rm e}/m_{\rm i})\sigma_{\rm e}$ and become significantly slower when $\sigma_{\rm i}<1$. This case will be studied separately. Here, as a first step, we consider the simpler regime $\sigma_{\rm i}>1$. 

As known from the simulations without cooling, the different particle masses become unimportant when $\sigma_{\rm i}\gg 1$ and reconnection proceeds very similarly to that in pair plasma, with nearly equal energy spectra of ions and electrons. Cooling, however, breaks this symmetry, because only electrons can be efficiently cooled while ions retain the energy acquired from field dissipation. Then, a two-temperature plasma forms. A key question is whether there is energy transfer from the hot ions to the cool electrons. It could provide gradual electron heating inside plasmoids, keeping them at a much higher temperature than found for reconnection in pair plasmas. Then, thermal electron motions could dominate over the bulk plasmoid motions, and thus also dominate the IC power of the reconnection layer.

In general, the electron-ion energy exchange may occur via Coulomb collisions (at sufficiently high densities) or via collisionless plasma processes. We defer a complete analysis of this problem to a future work and here use our PIC simulations to study only the collisionless energy exchange  (see also \citet{zhdankin_20b}, for a discussion of a similar problem in the context of plasma turbulence).

We devise a hybrid experiment mimicking a plasma composed of radiatively cooled electrons and hot ions. This experiment adopts the electron-positron composition, leveraging on the fact that reconnection in $\sigma_{\rm i}\gg1$ electron-ion plasmas is virtually indistinguishable from $\sigma\gg 1$ electron-positron reconnection \citep[e.g.,][]{guo_16b}, however we cool only one species (the electrons). This may be a useful first step for understanding collisionless energy exchange in
electron-ion radiative reconnection. 

Fig.~\ref{fig:2D_rec_layer_posicool_ion} shows the structure of the  reconnection layer in our hybrid experiment with $\sigma=10$. Electrons are cooled with radiation density that corresponds to $\gamma_{\rm cr,e^-}=16$ (same as the fiducial value of $\gcr$ in \citetalias{Sironi_20}). We observe that the two species have identical bulk motions (panels [c]), and significantly different internal motions (panels [d]). Positrons, which are not cooled, are especially hot inside plasmoids---the average Lorentz factor of positrons is comparable to $\sigma$, which is typical for reconnection models without cooling. By contrast, electrons are cooled to a non-relativistic temperature, and their energy is dominated by the plasmoid bulk motion. 

\begin{figure*}
\includegraphics[width=11cm]{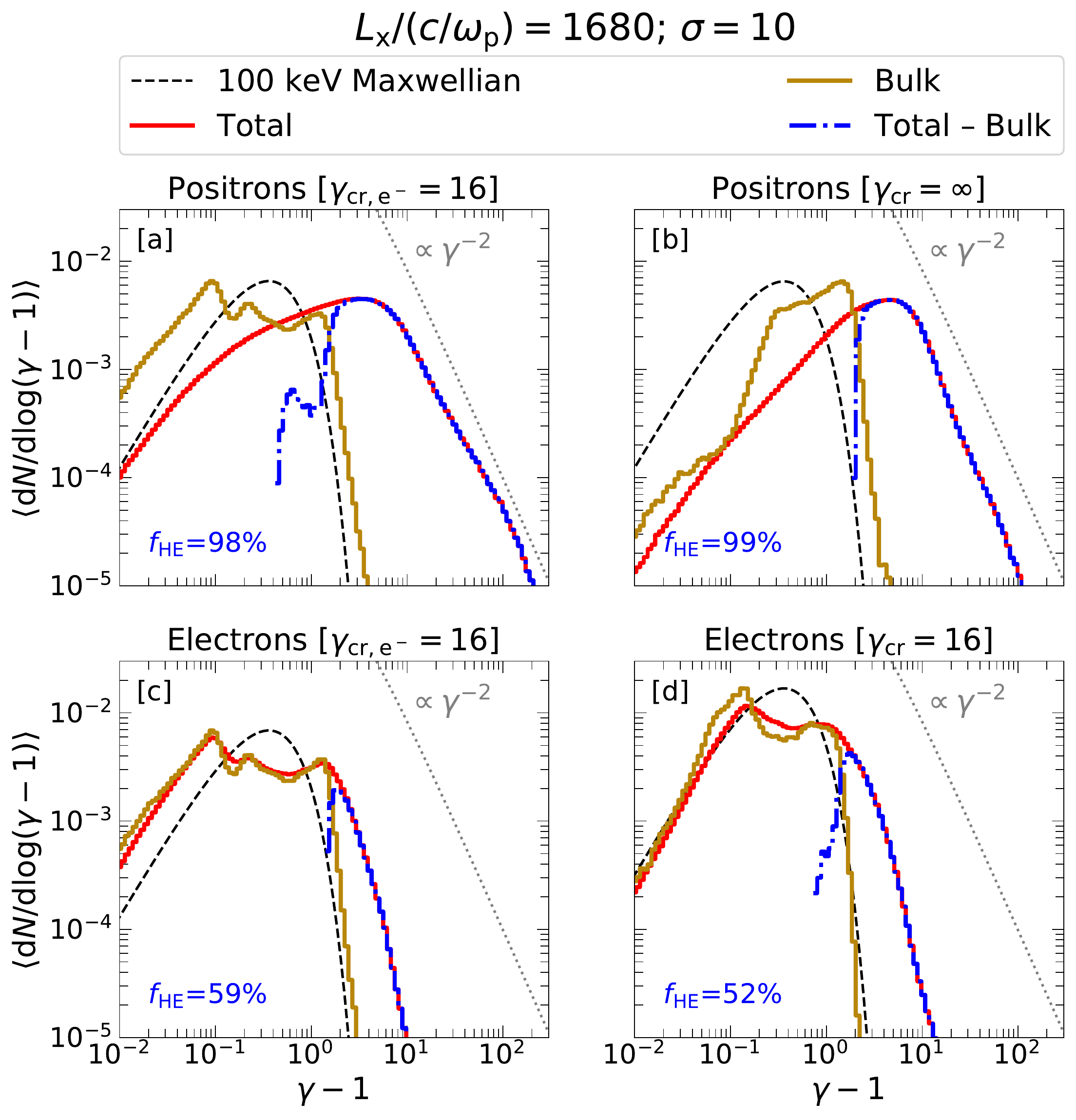}
\caption{
Particle energy spectra extracted from the reconnection region and averaged over time $2\lesssim T/(L_{\rm x}/c)\lesssim 5$ for three different simulations with magnetization $\sigma=10$ and box size $L_{\rm x}/(c/\omega_{\rm p})=1680$.
Two left panels show the results of our hybrid experiment (same simulation as in Fig.~\ref{fig:2D_rec_layer_posicool_ion}): [a] positrons and [c] electrons. For comparison, two other simulations are shown on the right: [b] model with no cooling ($\gcr=\infty$) and [d] model where both species are strongly cooled ($\gcr=16$).
In these two models, both species are treated equally and the $e^+/e^-$ spectra are identical. Each of the four panels shows  the total energy spectrum (red; $\gamma=\gamma_e$), the bulk motion spectrum (golden-brown; $\gamma=\Gamma$), and their difference (dash-dotted blue). Each panel also shows a Maxwellian distribution with temperature of 100 keV (dashed black), normalized so that its peak matches the peak of the bulk energy spectrum. The dotted grey lines indicate the $-2$ slope that corresponds to equal IC power per decade in Lorentz factor.
}
\label{fig:spectra_posicool}
\end{figure*}

Fig.~\ref{fig:spectra_posicool} shows the electron and positron spectra averaged over the reconnection region and time. The positron spectrum is similar to what was found in previous simulations with no cooling. It peaks at $\gamma_{\rm e}-1\sim 4$ and has a high-energy tail extending beyond $\gamma_{\rm e}\sim 10^2$. By contrast, the electron spectrum is similar to the simulations with full cooling (Section~\ref{subsec:spectra}). Electrons have ultra-relativistic internal motions only in the thin regions of the main reconnection layer or in the secondary layers in between merging plasmoids, where particles are being actively heated/accelerated.
In the hybrid experiment, these high-energy electrons carry a fraction $\fHE\sim59\%$ of the total IC power, remarkably similar to $\fHE\sim52\%$ found in a corresponding simulation where both species are cooled.

The main result of the hybrid simulation is that the behavior of electrons is nearly the same as in the simulation with full cooling of both $e^+$ and $e^-$. This demonstrates that the energy transfer from the hot positrons to the cool electrons is inefficient, in the sense that it is  unable to counteract the radiative losses of the electrons. This suggests that the electron component inside plasmoids will be kept cold also in electron-ion reconnection, at least when $\sigma_i\gg 1$. Then, Comptonization of radiation in the reconnection layer will remain dominated by the bulk motions of the plasmoid chain.

In the simulation that had strong cooling of both $e^+$ and $e^-$ we observed the formation of density cavities inside plasmoids (Section~\ref{subsec:general}). These cavities practically disappear in the hybrid simulation. This is because, the inertia of the quasi-neutral plasma inside plasmoids is now much higher (the effective mass per particle is $\sim \sigma m_{\rm e}$ because of the hot positrons) and therefore, the drag effect on the plasma distribution inside plasmoids is weaker.

\section{Monte Carlo radiative transfer calculations} \label{sec:radiative_transfer}

\begin{figure*} 
\includegraphics[width=13cm]{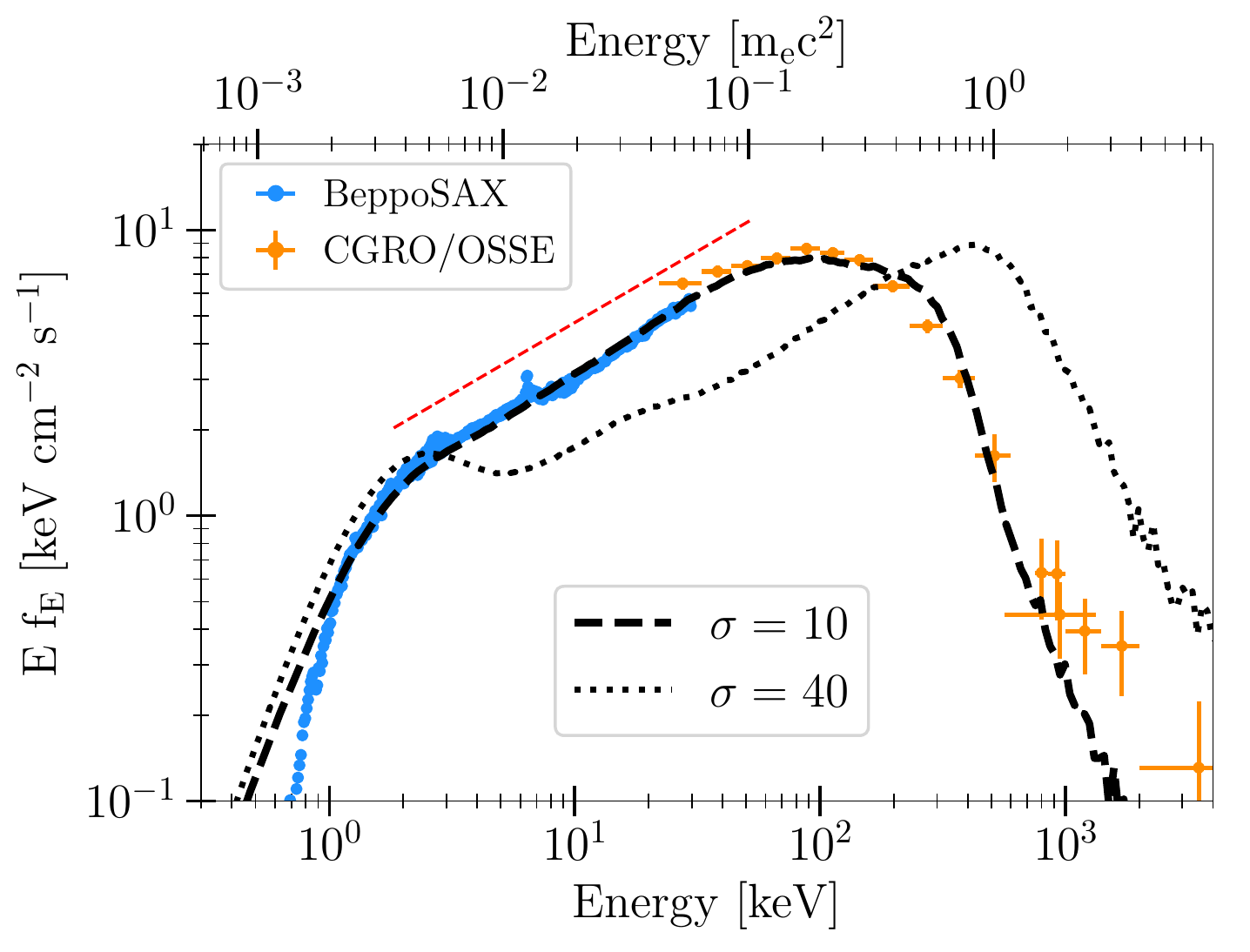}

\caption{X-ray/$\gamma$-ray spectrum of Cygnus~X-1 during its hard state, $Ef_{\rm E}=E^2N(E)$. The 0.7--25\,keV data (blue) are from \textit{BeppoSAX} \citep{DiSalvo+01, Frontera+01}, and the 25--3500\,keV data (orange) are from \textit{CGRO/OSSE} \citep{McConnell+02}. The black dashed and dotted curves show the spectra formed by Comptonization in the reconnection layer in the models with $\sigma=10$ (dashed) and $\sigma=40$ (dotted), with the same strength of radiative cooling losses ($\gcr=16$) and Compton amplification factor ($A=10$). The dashed red line indicates the power-law $N(E)\propto E^{-1.5}$. All the data are normalized with respect to \textit{OSSE}.}
\label{fig:X-ray_spectra}
\end{figure*}

We perform Monte-Carlo simulations of X-ray emission expected from the reconnection layer using the radiative transfer code \texttt{CompPair} \citep{Beloborodov_20}. In the simulation, seed soft radiation is injected in the mid-plane of the reconnection layer, with a Planck spectrum of temperature $kT_s=10^{-3}m_{\rm e}c^2$, and the code follows the scattering of photons in the reconnection region, leading to formation of a Comptonized, hard X-ray spectrum. The radiative transfer is calculated in the region of $|x|<H=L_{\rm x}/3$, where most of the scattering is expected to occur in a self-consistently produced $e^\pm$ plasma \citep{belo_17}.

We use here a simplified setup of the transfer simulation which does not calculate $e^\pm$ production self-consistently. The inflow density $n_{\rm 0}$ is fixed in our PIC simulations, with no pair annihilation or injection of new $e^\pm$ pairs. We use 7 snapshots of the PIC simulations spaced with time intervals of $0.5L_{\rm x}/c$ between $T/(L_{\rm x}/c)=2$ and $T/(L_{\rm x}/c)=5$. The region $|x|<L_{\rm x}$, $|y|<H$ is divided into $60\times 20$ patches, and for each scattering event, the particle momentum is drawn from the local distribution function in the patch. The transfer simulation is then evolved in time with updating the distribution function every $0.5L_{\rm x}/c$. The result of the radiative transfer depends on the overall normalization of the plasma density, which we parameterize using Thomson optical depth $\tau_{\rm T}=H\sigma_{\rm T}n_{\rm 0}$, defined in terms of the inflow density $n_{\rm 0}$. The optical depth is expected to be $\tau_{\rm T}\sim 1$, depending on the compactness of the magnetic flare. Therefore, the photons sample the particle population on the photon mean free path to scattering, which is not smaller than $H$. We adjust $\tau_{\rm T}$ to achieve the Compton amplification factor\footnote{The factor $A$ is the ratio of the average energy of escaping photons to the average energy of injected soft photons.} $A\approx 10$, which is typically required for reproducing observed hard state spectra of accreting black holes, with a photon index of $\Gamma\sim1.5$ \citep{belo_17}. The corresponding optical depth is $\tau_{\rm T}=1.5$ for $\sigma=10$, and $\tau_{\rm T}=0.5$ for $\sigma=40$.

Fig.~\ref{fig:X-ray_spectra} shows the spectrum of radiation escaping from the reconnection region for the two models, with $\sigma=10$ and $\sigma=40$, adopting the same $\gamma_{\rm cr}=16$. The emitted spectrum varies with time as the reconnection proceeds, and Fig.~\ref{fig:X-ray_spectra} shows the emission averaged over the time interval $2\lesssim T/(L_{\rm x}/c)\lesssim5$. We observe that the spectrum peaks
around 100\,keV and has a steeply decaying tail at higher energies. The model with $\sigma=10$ is remarkably close to the typical hard-state of Cygnus~X-1 observed in a broad band of $1-1000$\,keV. For comparison, we over-plot in Fig.~\ref{fig:X-ray_spectra} the data available from \textit{BeppoSAX} and \textit{CGRO/OSSE}. The deviation of the data from the model at low photon energies ($\lesssim 1$\,keV) is expected from inter-stellar absorption. The excess at higher energies ($\gtrsim$1\,MeV) may be interpreted as a contribution from coronal regions with $\sigma>10$. Our simulations demonstrate that with increasing $\sigma$ the Comptonized spectrum develops a stronger high-energy tail (compare the models with $\sigma=10$ and $\sigma=40$ in Fig.~\ref{fig:X-ray_spectra}). The increased $\sigma$ also shifts the spectral peak to above 100~keV. This shift is caused by the combination of two effects: the increased dispersion of the plasmoid speeds at high $\sigma$ (Fig.~\ref{fig:std_ux}), and the fact that high-energy particles contribute a larger fraction of IC power  at higher magnetizations (Fig.~\ref{fig:f_HE}).

\section{Summary} \label{sec:conclusion}

In this paper, we have performed 2D PIC simulations of relativistic magnetic reconnection with strong IC cooling, for conditions appropriate to the strongly magnetized regions of black hole coronae. Our main goal was to explore whether Comptonization by the Compton-cooled plasmoid chain in the reconnection layer can explain the observed hard X-ray spectra of accreting black holes, as proposed by \citet{belo_17}. Our work extends the investigation of \citetalias{Sironi_20} to higher magnetizations up to $\sigma=40$ and includes the Monte-Carlo radiative transfer simulations of Comptonization in the reconnection region using the particle momentum distribution directly derived from the PIC simulations. 
Our main results are as follows.

\begin{enumerate}
\item 
For a given magnetization $\sigma$, the energetics of the plasma in the reconnection layer is uniquely determined by the parameter $\gcool=t_{\rm cool}/t_{\rm adv}$, where $t_{\rm cool}$ is the timescale for particles to cool to a non-relativistic  energy (due to IC scattering), and $t_{\rm adv}\approx L_{\rm x}/c$ is the advection timescale along the layer of size $L_{\rm x}$. In the relevant limit of $\gcool\ll 1$, the majority of particles found inside the plasmoids are cooled to a non-relativistic internal energy per particle $\sim 0.5 \gcool m_{\rm e}c^2$ (if it still exceeds the Compton temperature of the radiation field, $kT_{\rm C}\ll m_{\rm e}c^2$). This makes thermal Comptonization unable to generate 100 keV X-rays.

\item 
In the strongly cooled regime, the IC power output is dominated by the bulk motions of a cold-chain of plasmoids, which are pulled against Compton drag by the magnetic stresses in the reconnection layer. The mean bulk 4-velocity along the layer is nearly independent of the cooling strength $\gcool^{-1}$ or the magnetization $\sigma\gg 1$, while the dispersion of bulk motions around the mean value tends to increase with increasing magnetization and $\gcool$ (i.e., for weaker cooling).

\item Particles accelerated in the reconnection layer on a timescale faster than the IC losses form a high-energy tail in the particle spectrum. They are localized in thin regions of the main layer or of secondary layers between merging plasmoids. In the limit of strong cooling, we find that the fraction of radiated IC power contributed by the high-energy tail is $\sim 25$\% for $\sigma=10$ and $\sim 40$\% for $\sigma=40$.

\item 
Comptonization of seed soft photons in the reconnection layer with $\sigma=10$ closely reproduces the typical X-ray spectrum observed in the hard state of accreting black holes (e.g., with a power-law index of photon flux density $\lesssim1.8$). It has a peak near 100~keV, which is shaped by the bulk motions of cold plasmoids. At higher $\sigma$ the peak shifts to higher energies and the  MeV tail of the spectrum increases.
\end{enumerate}

Our simulations adopt an electron-positron plasma, and we defer a study of radiative reconnection in electron-ion plasmas to Part II of this series. However, in the current work, we have also presented a strategy for mimicking a plasma composed of radiatively cooled electrons and hot ions, by performing a hybrid experiment in which we still adopt an electron-positron composition, but cool only one species (the electrons). Then, the positrons play the role of ions (which do not suffer radiative losses) with a reduced mass $m_{\rm i}=m_{\rm e}$. This experiment is likely to bear similarities to the full electron-ion model with $m_{\rm i}\gg m_{\rm e}$, in view of the known similarity of electron-positron and electron-ion relativistic reconnection in the regime of $\sigma\gg1$ and negligible cooling.

Our hybrid experiment shows that the two species do not equilibrate, i.e., energy transfer from the uncooled positrons to the cooled electrons is not sufficiently fast to offset the IC losses of the electrons. As a result, the electron energies remain dominated by the bulk motions in the plasmoid chain. Thus, the cold-chain Comptonization model may also apply to radiative relativistic reconnection in electron-ion plasmas.

Our simulations are two-dimensional, and we refer to \citetalias{Sironi_20} for a demonstration of the fact that radiative 3D simulations lead to the same conclusions as their 2D counterparts.
We refer to Part II of this series for an investigation of radiative reconnection in electron-ion plasmas, in the trans-relativistic regime $\sigma\sim1$. In the future, it may also be useful to investigate how the results change in the presence of a strong guide field, comparable to the reconnecting fields.

\section*{Acknowledgements}

This paper benefited from useful discussions with Riley M.~T.~Connors, Javier A.~Garc\'{i}a, Victoria Grinberg, Guglielmo Mastroserio and James F.~Steiner. We thank Andrzej Zdziarski for sharing details pertaining to the hard state observations of Cygnus~X-1. N.S. acknowledges the support from Columbia University Dean's fellowship. L.S. acknowledges support from the Sloan Fellowship, the Cottrell Scholars Award, NASA 80NSSC20K1556, NSF PHY-1903412, and DoE DE-SC0021254. A.M.B. is supported by NSF grants AST-1816484 and AST-2009453, Simons Foundation grant \#446228, and the Humboldt Foundation. This project made use of the following computational resources: Habanero and Terremoto HPC clusters at Columbia University, and Cori of NERSC.

\section*{Data availability}
The data underlying this article will be shared on reasonable request to the authors.

%%%%%%%%%%%%%%%%%%%%%%%%%%%%%%%%%%%%%%%%%%%%%%%%%%

%%%%%%%%%%%%%%%%%%%% REFERENCES %%%%%%%%%%%%%%%%%%

% The best way to enter references is to use BibTeX:
\bibliographystyle{mnras}
\bibliography{main} % if your bibtex file is called example.bib

%%%%%%%%%%%%%%%%%%%%%%%%%%%%%%%%%%%%%%%%%%%%%%%%%%

%%%%%%%%%%%%%%%%% APPENDICES %%%%%%%%%%%%%%%%%%%%%

\appendix

\section{Simulation parameters}\label{appendix:table}
We present in Table \ref{tab:setup} the input parameters of our simulations (left side) and the main outcomes (right side). 

\begin{table*}
\setlength{\tabcolsep}{3pt}
\begin{center}
      \caption{List of numerical input parameters corresponding to different simulations, and the resulting plasma properties.}
      \label{tab:setup}
      \begin{tabular}{ccccc| @{\vline} |ccccc}
       \hline
       \hline
      $\sigma^{[1]}$ & $\gamma_{\rm cr}^{[2]}$ & \multicolumn{2}{c}{Domain size$^{[3]}$} & $\gcool^{[4]}$ & $\langle \Eint  \rangle^{[5]}$ & $\langle \Gamma \rangle - 1^{[6]}$ & $\langle u_{\rm x} (x>0) \rangle_{\rm x}^{[7]}$ & $\langle \sigma_{u_{\rm x}} \rangle_{\rm x}^{[8]}$ & $f_{\rm HE}^{[9]}$\\ 
      &  & $L_{\rm x}/(c/\omega_{\rm p})^{[a]}$ & $L_{\rm x}/(r_{\rm 0,hot})^{[b]}$ & & & & & & (\%) \\ % <--
       \hline
       \hline
         \multicolumn{10}{c}{Very strongly magnetized regime} \\
       \hline
      40 & 16   & 840 & 132.8 & 0.48 &  0.46 &  0.38 & 0.98 & 0.76 &  77 \\ % <--
      40 & 22.6   & 840 & 132.8 & 0.96 &  0.75 &  0.30 & 0.80 & 0.77 &  89 \\ % <--
      40 & 32   & 840 & 132.8 & 1.92 &  1.18 &  0.38 & 0.72 & 0.76 &  92 \\ % <--
      40 & 45   & 840 & 132.8 & 3.85 &  1.71 &  0.36 & 0.78 & 0.68 &  96 \\ % <--
      40 & $\infty$ & 840 & 132.8 & $\infty$ & 7.52 & 0.26 & 0.56 & 0.76 & 100 \\ % <--
       \hline
      40 & 16  & 1680 & 265.6 & 0.24 &  0.19 &  0.40 & 1.04 & 0.72 &  59 \\ % <--
      40 & 22.6  & 1680 & 265.6 & 0.48 &  0.39 &  0.39 & 0.94 & 0.82 &  78 \\ % <--
      40 & 32  & 1680 & 265.6 & 0.96 &  0.64 &  0.39 & 0.86 & 0.82 &  88 \\ % <--
      40 & 45  & 1680 & 265.6 & 1.92 &  1.21 &  0.41 & 0.88 & 0.82 &  93 \\ % <--
      40 & $\infty$  & 1680 & 265.6 & $\infty$ & 7.20 & 0.44 & 1.08 & 0.96 & 100 \\ % <--
       \hline
      40 & 16   & 3360 & 531.2 & 0.12 &  0.13 &  0.35 &  0.92 &  0.60 &  51\\ % <--
      40 & 22.6  & 3360 & 531.2 & 0.24 &  0.20 &  0.36 &  1.04 &  0.80 &  64\\ % <--
      40 & 32  & 3360 & 531.2 & 0.48 &  0.39 &  0.42 &  1.10 &  0.82 &  77\\ % <--
      40 & 45  & 3360 & 531.2 & 0.96 &  0.76 &  0.36 &  0.92 &  0.82 &  89\\ % <--
      40 & $\infty$  & 3360 & 531.2 & $\infty$ & 8.29 & 0.41 & 0.93 & 0.86 & 100 \\ % <--
       \hline
      40 & 16  & 6720 & 1062.4 & 0.06 &  0.06 & 0.43 &  0.99 &  0.69  & 39\\ % <--
      \hline
      \hline
         \multicolumn{10}{c}{Strongly magnetized regime} \\
       \hline
      10 & 11.3  & 420 & 132.8 & 0.96 &  0.35 &  0.36 &  0.48 &  0.25 &  71 \\ % <--
       \hline
      10 & 8  & 840 & 265.6 & 0.24 & 0.09 & 0.17 & 0.58 & 0.34 & 44 \\ % <--
      10 & 11.3   & 840 & 265.6 & 0.48 & 0.19 & 0.32 & 0.80 & 0.44 & 58 \\ % <--
      10 & 16   & 840 & 265.6 & 0.96 & 0.32 & 0.43 & 1.02 & 0.56 & 71 \\ % <--
      10 & 22.6   & 840 & 265.6 & 1.92 & 0.53 & 0.46 & 1.04 & 0.58 & 84 \\ % <--
      10 & 32   & 840 & 265.6 & 3.84 &  0.82 &  0.51 & 1.28 & 0.62 &  89 \\ % <--
       \hline
      10 & 8  & 1680 & 531.2 & 0.12 & 0.03 & 0.13 & 0.58 & 0.30 & 34 \\ % <--
      10 & 11.3  & 1680 & 531.2 & 0.24 & 0.08 & 0.22 & 0.82 & 0.48 & 46 \\ % <--
      10 & 16  & 1680 & 531.2 & 0.48 &  0.15 &  0.27 &  0.84 &  0.52 & 52 \\ % <--
      10 & 16$^\dagger$  & 1680 & 531.2 & 0.48 & 0.09 & 0.20 & 0.72 & 0.50 & 59 \\% <--
      10 & 22.6  & 1680 & 531.2 & 0.96 & 0.39 & 0.40 & 1.02 & 0.58 & 77 \\ % <--
      10 & 32  & 1680 & 531.2 & 1.92 &  0.64 &  0.50 & 1.20 & 0.62 &  87 \\ % <--
       \hline
      10 & 8  &  3360 & 1062.4 & 0.06 & 0.01 & 0.08 & 0.52 & 0.20 & 23 \\ % <--
      10 & 11.3   & 3360 & 1062.4 & 0.12 & 0.03 & 0.16 & 0.68 & 0.30 & 28 \\ % <--
      10 & 16   & 3360 & 1062.4 & 0.24 &  0.07 &  0.37 &  1.06 &  0.48 &  39  \\ % <--
       \hline
      10 & 11.3   & 6720 & 2124.8 & 0.06 &  0.02 &  0.12 & 0.68 & 0.26 &  25 \\ % <--
      10 & 16   & 6720 & 2124.8 & 0.12 &  0.03 &  0.24 &  0.90 &  0.36 &  30 \\ % <--
	 \hline
    \end{tabular}
      \begin{tablenotes}
      \footnotesize
      \item \textit{Note.} {The first five columns contain numerical and physical input parameters, and the last five columns contain the plasma parameters obtained from the simulations. All simulations are performed for a duration of $T\sim5L_{\rm x}/c$. The description of each column is as follows. $^{[1]}$Magnetization in the upstream plasma; $^{[2]}$Critical Lorentz factor---a proxy for the intensity of incident photon field ($\gamma_{\rm cr}=\infty$ implies no IC cooling, and smaller $\gamma_{\rm cr}$ implies greater IC cooling; see Eq.~\ref{eqn:gamma_cr});  $^{[3]}$Half-length of the computational domain along the $x$-direction in units of $^{[a]}$plasma skin depth, and $^{[b]}$post-reconnection Larmor radius $r_{\rm 0,hot}=\sqrt{\sigma}\comp$, respectively; $^{[4]}$ $\gcool$, as defined in \eq{varrho}; $^{[5,6]}$mean internal and bulk energy per particle in units of rest mass energy ($m_{\rm e}c^2=511$~keV); $^{[7]}$bulk outflow dimensionless 4-velocity along the reconnection layer averaged over one half of the box; $^{[8]}$spread of the 4-velocity distribution averaged over one half of the box; $^{[9]}$fractional IC power contributed by high-energy particles. The symbol $^\dagger$ indicates the hybrid experiment (see \S\ref{subsec:posicool}) where only electrons are subjected to IC cooling while positrons are not, i.e., $\gamma_{\rm cr,e^-}=16$ and $\gamma_{\rm cr,e^+}=\infty$.}
      \end{tablenotes}
\end{center}
\end{table*}

\section{Plasmoid properties} \label{appendix:plasprop}

We identify plasmoids based on the $z$-component of the magnetic vector potential ($A_{\rm z}$), as described in \citet{sironi_16}. We first identify O-points as maxima of  $A_{\rm z}$, which correspond to the plasmoid centers. Along the reconnection layer, the two local minima around each O-point correspond to two X-points. The largest of the two values of  $A_{\rm z}$ at these two X-points identifies the equipotential line that we use as the plasmoid contour. 

\begin{figure} 
\includegraphics[width=8cm]{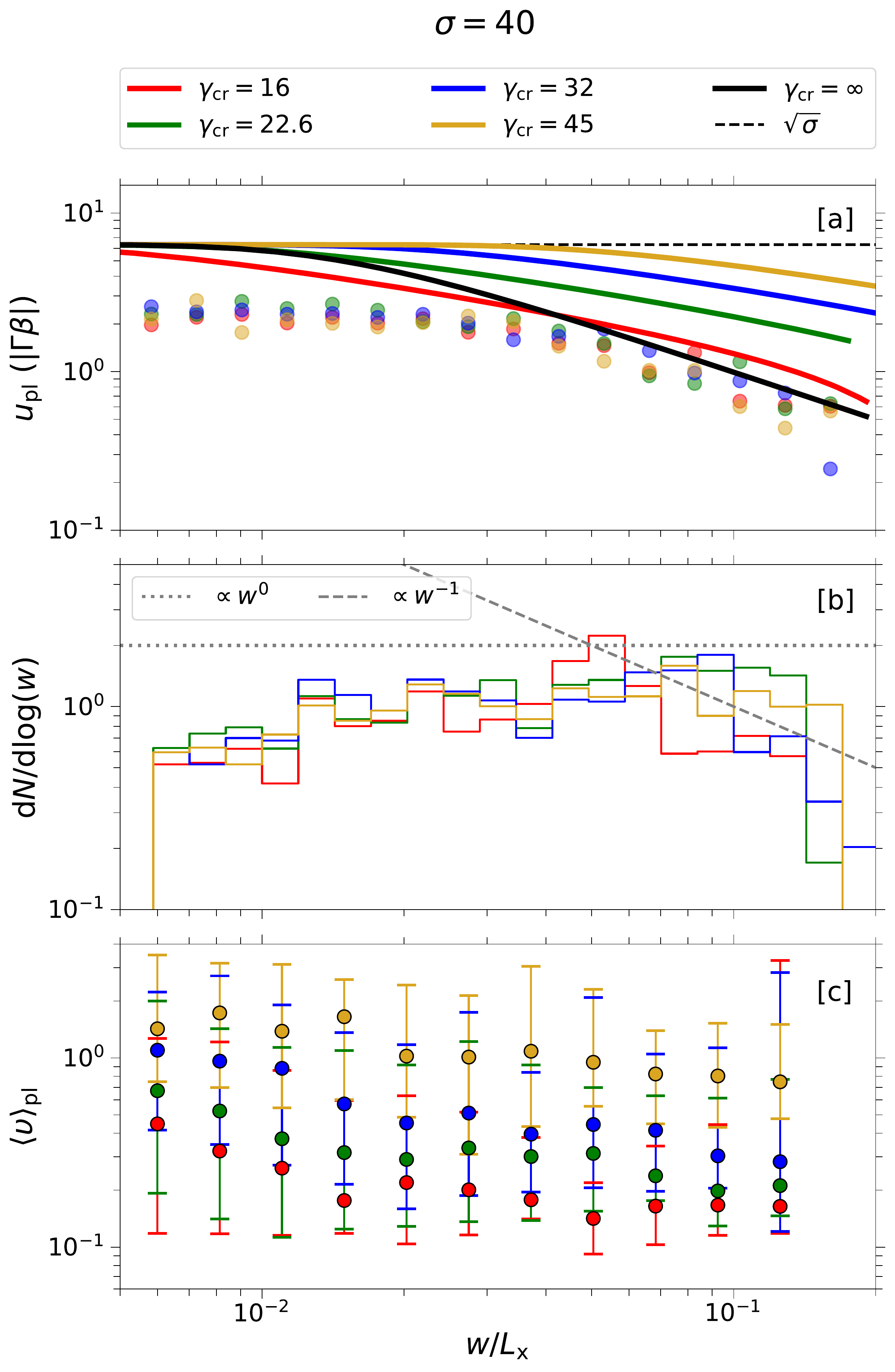}
    \caption{Plasmoid properties in a strongly magnetized ($\sigma=40$) plasma, as a function of their width $w/L_{\rm x}$ and of the cooling strength $\gamma_{\rm cr}$ (red: $\gamma_{\rm cr}=16$, green: $\gamma_{\rm cr}=22.6$, blue: $\gamma_{\rm cr}=32$, golden brown: $\gamma_{\rm cr}=45$). The simulations are performed with a $L_{\rm x}/(c/\omega_{\rm p})=3360$ box, and each of the quantities displayed is averaged over $2\lesssim T/(L_{\rm x}/c)\lesssim5$. 
    \textit{Top panel [a]}: Each circular marker denotes the 90$^{\rm th}$ percentile of plasmoid dimensionless 4-velocities for plasmoids of a given size. The solid black and colored lines indicate the upper limit in the absence (Eq.~\ref{eq:bulk2}) and presence (Eq.~\ref{eq:outflow_drag}) of IC cooling, respectively. For comparison, the Alfv\'en limit is shown by the dashed horizontal line.
    \textit{Middle panel [b]}: Cumulative distribution of plasmoid widths $F(w)={\rm d}N/{\rm d\log{w}}$. The expectations $F(w)\propto w^{0}$ \protect\citep{huang_12} and $F(w)\propto w^{-1}$\protect\citep{uzdensky_10,loureiro_12} are depicted by the grey dotted and dashed lines, respectively. 
    \textit{Bottom panel [c]}: Internal energy per particle (in units of $m_{\rm e}c^2=511$~keV), averaged over all the particles inside a plasmoid. The filled circles denote the median values for a given plasmoid width, and the error bars denote the 10$^{\rm th}$ and 90$^{\rm th}$ percentiles.}
\label{fig:plasprop}
\end{figure}

We define the plasmoid width $w$ in the direction transverse to the main layer, and the plasmoid bulk velocity as the velocity of its O-point (plasmoids move nearly as rigid bodies). In \fig{plasprop}, we describe the properties of plasmoids as a function of their width $w$, for different levels of IC cooling. Panel [a] quantifies the plasmoid maximum bulk 4-velocity (more precisely, circles indicate the values of the 90$^{\rm th}$ percentile in each size bin). In the absence of IC cooling, the plasmoid 4-velocity when exiting the layer\footnote{For plasmoids that do not end their life by merging into a bigger plasmoid, this corresponds to the time when they are largest and fastest.} is
\be\label{eq:bulk2}
\Gamma\beta \simeq \sqrt{\sigma} \tanh\left(\frac{\eta_{\rm rec}}{\sqrt{\sigma}}\frac{L_{\rm x}}{w}\right)
\ee
which is indicated by the black line in panel [a]. When strong Compton drag is included, the plasmoid bulk motion is regulated by the competition between bulk acceleration by magnetic field tension and Compton drag. As shown by \citetalias{Sironi_20}, this leads to a maximum 4-velocity
\begin{equation}  \label{eq:outflow_drag}
\Gamma\beta \simeq  \sqrt{\sigma}\tanh\bigg(\frac{2n_{\rm 0}}{n'_{\rm pl}}\frac{c/\omega_{\rm p}}{w}\frac{\gamma_{\rm cr}^2\beta}{\Gamma}\bigg),
\end{equation}
where $n'_{\rm pl}/n_{\rm 0}$ is the plasmoid compression factor, defined as the ratio of the mean rest-frame density in the plasmoid to the upstream density. The curves corresponding to different levels of cooling are shown in panel [a] by the different colors (see legend). At each value of $w$, the maximum 4-velocity of a plasmoid will be controlled by the smallest between \eq{bulk2} and \eq{outflow_drag}. The data points shown in panel [a] demonstrate that Compton drag does not appreciably slow down the plasmoids in our simulations. Also, regardless of the cooling level, smaller plasmoids are faster than larger plasmoids, yet they barely reach the Alf\'enic limit $\Gamma\beta \sim \sqrt{\sigma}\simeq 6.3$ predicted by \citet{lyubarsky_05} (indicated by the horizontal dashed line).

In the middle panel [b] of Fig.~\ref{fig:plasprop}, we present the cumulative plasmoid size distribution $F(w)={\rm d}N/{\rm d}\log(w)$. Regardless of the level of cooling, this can be modeled as a broken power-law, with $F(w)\propto w^{0}$ \citep{huang_12} for small sizes and $F(w)\propto w^{-1}$ \citep{uzdensky_10, loureiro_12} for larger plasmoids, $w/L_{\rm x}\gtrsim0.07$.

The bottom panel [c] of Fig.~\ref{fig:plasprop} shows the mean internal energy per particle inside plasmoids of different sizes. We generally find that larger plasmoids, which live longer, have time to cool down to lower temperatures. This effect is less pronounced for moderate cooling (yellow points for $\gcr=45$; here, all plasmoids stay rather hot), and more evident for stronger cooling. As expected, for plasmoids of a given size, the average internal energy per particle decreases for stronger cooling.

\section{Computation of the internal energy per particle} \label{appendix:internal}

The probability distribution function of a plasma described by a Maxwell-J\"{u}ttner distribution is
\begin{equation} \label{eqn:max-jut}
f_{\rm MJ}(\gamma_{\rm e},\theta_{\rm e}) = \frac{\gamma_{\rm e}\sqrt{\gamma_{\rm e}^2 -1}\exp{(-\gamma_{\rm e}/\theta_{\rm e})}}{\theta_{\rm e}K_2\left({\frac{1}{\theta_{\rm e}}}\right)},
\end{equation}
where $\theta_{\rm e} = kT_{\rm e}/m_{\rm e} c^2$ is the dimensionless temperature parameter and $K_2$ is the modified Bessel function of the second kind. The internal energy per particle (in units of $m_{\rm e} c^2$) can be written as
\begin{equation} \label{eqn:int_en_ppl}
\Eint  = \dfrac{\int_{1}^{\infty}{\gamma_{\rm e} f_{\rm MJ}(\gamma_{\rm e},\theta_{\rm e})}d\gamma_{\rm e}}{\int_{1}^{\infty}{f_{\rm MJ}(\gamma_{\rm e},\theta_{\rm e})}d\gamma_{\rm e}} -1.
\end{equation}
For values of $\theta_{\rm e}$ in the range [$10^{-3},10^{3}$], Eq.~\ref{eqn:int_en_ppl} is solved for $\Eint  (\theta_{\rm e})$. This can be used to find a mapping between the adiabatic index $\hat{\gamma_{\rm e}}(\Eint )$ and $\Eint $ using the relation $\theta_{\rm e}=[\hat\gamma_{\rm e}(\Eint )-1]\Eint $. We then model this mapping by assuming that the adiabatic index is of the form 
\begin{equation} \label{eqn:ad_index_abcd}
\hat\gamma_{\rm e}\equiv\dfrac{A+B\Eint }{C+D\Eint },
\end{equation}
where the best fit values of the numerical coefficients are found to be $A\simeq1.187$, $B\simeq1.251$, $C\simeq0.714$, and $D\simeq0.936$. Note that the numerical coefficients satisfy $A/C\simeq5/3$ and $B/D\simeq4/3$ in the non-relativistic ($\Eint \rightarrow0$) and ultra-relativistic  ($\Eint \rightarrow\infty$) limits, respectively. 

In order to estimate the internal energy per particle ($\Eint $) and the temperature ($\theta_{\rm e}$) in the fluid frame, we assume a perfect and isotropic fluid for which the stress-energy tensor is given by
\begin{equation} \label{eqn:stress-energy}
T^{\mu\nu}=(e+p)U^{\mu}U^{\nu}-pg^{\mu\nu},
\end{equation}
where $p$ is the pressure, $U^{\mu}$ is the fluid dimensionless four-velocity, $g^{\mu\nu}$ is the flat-space Minkowski metric, and $e=n'_{\rm e}m_{\rm e}c^2+u_{\rm e}$ is the rest-frame energy density of electrons, where $n_{\rm e}'$ is the particle number density in the fluid frame, and $u_{\rm e}=p_{\rm e}/(\hat{\gamma}-1)$ is the internal energy density (like electron pressure $p_{\rm e}$, it is also defined in the fluid frame). The dimensionless internal energy per particle in the fluid frame is $\Eint=u_e/n_e'm_ec^2$. Using the transformation of $T^{\mu\nu}$ one can express $\Eint$ in terms of the lab-frame quantities,
\begin{equation}
\Eint  = \dfrac{[T_{\rm e}^{00}/n_{\rm e}m_{\rm e}c^2 - \Gamma]\Gamma}{1+\hat\gamma_{\rm e}(\Gamma^2-1)}
= \dfrac{[\bar{\gamma}_{\rm e} - \Gamma]\Gamma}{1+\hat\gamma_{\rm e}(\Gamma^2-1)},
\end{equation}
where $\Gamma$ is the fluid Lorentz factor, $\bar{\gamma}_{\rm e}$ is the average particle Lorentz factor, $T_{\rm e}^{00}=\bar{\gamma}_{\rm e} n_{\rm e} m_{\rm e}c^2$ is the energy density, and $n_{\rm e}=\Gamma n'_{\rm e}$ is the number density,  all measured in the lab frame.

\section{Assessment of quasi-steady state} \label{appendix:time_convergence}

Many of the results presented in this paper are obtained by averaging in the interval $2 \lesssim T/(L_{\rm x}/c) \lesssim 5$. As we describe in \S \ref{sec:setup}, at even earlier times ($T/(L_{\rm x}/c) \lesssim 2$), the reconnection layer is not in steady state---this is the time needed for the two reconnection fronts to advect out of the box the hot plasma initialized in the current sheet.

\begin{figure}
\includegraphics[width=8cm]{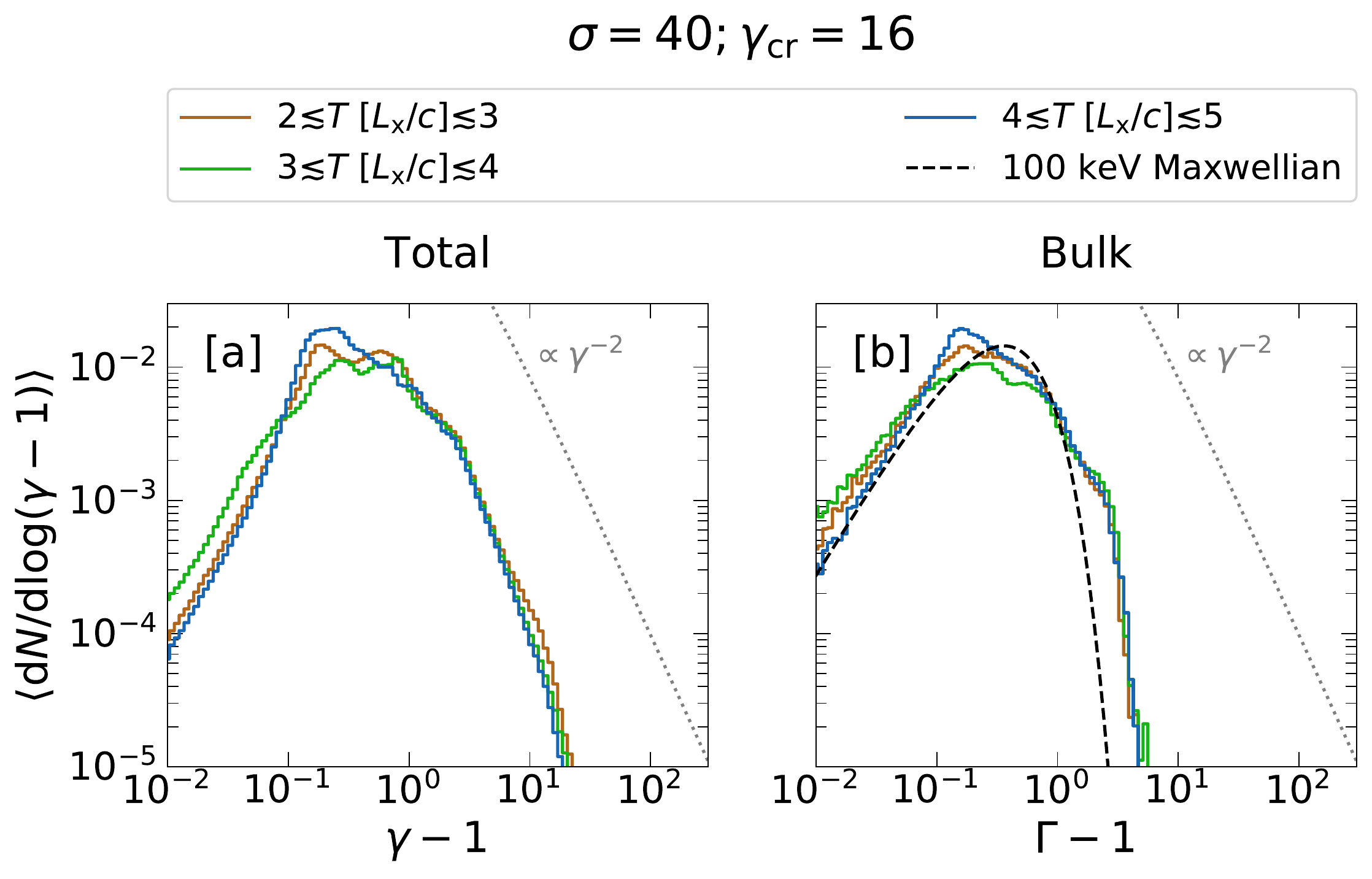}
\caption{Total (panel [a]) and bulk (panel [b]) energy spectra, time-averaged in the intervals $2\lesssim T/(L_{\rm x}/c)\lesssim 3$ (golden-brown), $3\lesssim T/(L_{\rm x}/c)\lesssim 4$ (green), and $4\lesssim T/(L_{\rm x}/c)\lesssim 5$ (blue). We employ our reference simulation with  $\sigma=40$, $\gamma_{\rm cr}=16$ and $L_{\rm x}/(c/\omega_{\rm p})=3360$. Panel [b] also shows a representative 100 keV Maxwellian (dashed black curve). Both panels demonstrate that the system achieves a quasi-steady state for $T/(L_{\rm x}/c)\gtrsim 2
$.}
\label{fig:spectra_time_convergence}
\end{figure}

In Fig.~\ref{fig:spectra_time_convergence}, we sub-divide the range $2 \lesssim T/(L_{\rm x}/c) \lesssim 5$ in intervals of duration $L_{\rm x}/c$, and in each interval we compute the time-averaged total [left] and bulk [right] spectrum. We employ our reference simulation with $\sigma=40$, $\gamma_{\rm cr}=16$ and $L_{\rm x}/(c/\omega_{\rm p})=3360$, whose spectrum is shown in Fig.~\ref{fig:spectra_domain}[c]. The figure convincingly demonstrates that the layer has achieved a quasi-steady state.

% Don't change these lines
\bsp	% typesetting comment
\label{lastpage}
\end{document}